\documentclass[sigconf]{acmart}
\AtBeginDocument{%
  \providecommand\BibTeX{{%
    \normalfont B\kern-0.5em{\scshape i\kern-0.25em b}\kern-0.8em\TeX}}}

\setcopyright{acmlicensed}
\copyrightyear{2026}
\acmYear{2026}
\acmDOI{https://doi.org/10.1145/3830398.3830663}

\acmConference[UIST '26]{The 39th Annual ACM Symposium on User Interface Software
and Technology (UIST ’26)}{November 02--05,
  2026}{Detroit, MI, USA}
\acmBooktitle{The 39th Annual ACM Symposium on User Interface Software
and Technology (UIST ’26),
 November 02--05, 2026, Detroit, MI, USA} 
\acmISBN{979-8-4007-2856-3/26/11}

\usepackage{multirow}
\usepackage{framed,enumitem} 
\usepackage{xspace}
\usepackage{makecell}
\usepackage{soul}
\usepackage{textcomp}
\usepackage{array}

\definecolor{mygreen}{RGB}{0,150,0}
\definecolor{myblue}{RGB}{30,100,200}
\definecolor{myred}{RGB}{180,0,0}
\definecolor{myyellow}{RGB}{200,150,0}
\setulcolor{mygreen}                
\setul{0.3ex}{0.3ex}    

\newcommand{\sysname}{\textsc{Contexty}}

\definecolor{blockcolor}{HTML}{000000}
\definecolor{blockrule}{gray}{0.6}

\begin{document}


\title[\sysname{}]{\sysname{}: Capturing and Organizing In-situ Thoughts for Context-Aware AI Support}

\author{Yoonsu Kim}
\orcid{0000-0002-9782-086X}
\affiliation{\institution{School of Computing \\ KAIST}
\city{Daejeon}
\country{Republic of Korea}}
\email{yoonsu16@kaist.ac.kr}

\author{Chanbin Park}
\orcid{0009-0009-0194-3995}
\affiliation{
\institution{University of California, Berkeley}
\city{Berkeley, CA}
\country{USA}
}
\email{chanbin.park@berkeley.edu}

\author{Kihoon Son}
\orcid{0000-0001-7224-2947}
\affiliation{\institution{School of Computing \\ KAIST}
\city{Daejeon}
\country{Republic of Korea}}
\email{kihoon.son@kaist.ac.kr}

\author{Saelyne Yang}
\orcid{0000-0003-1776-4712}
\affiliation{\institution{Carnegie Mellon University}
\city{Pittsburgh, PA}
\country{USA}}
\email{saelyney@andrew.cmu.edu}

\author{Juho Kim}
\email{juhokim@kaist.ac.kr}
\orcid{0000-0001-6348-4127}
\affiliation{
    \institution{School of Computing, KAIST}
    \city{Daejeon}
    \country{Republic of Korea}
}
\email{juho@skillbench.com}
\affiliation{
    \institution{SkillBench}
    \city{Santa Barbara, CA}
    \country{USA}
}

\renewcommand{\shortauthors}{Yoonsu Kim et al.}



\begin{teaserfigure}
  \centering
  \includegraphics[width=1\textwidth]{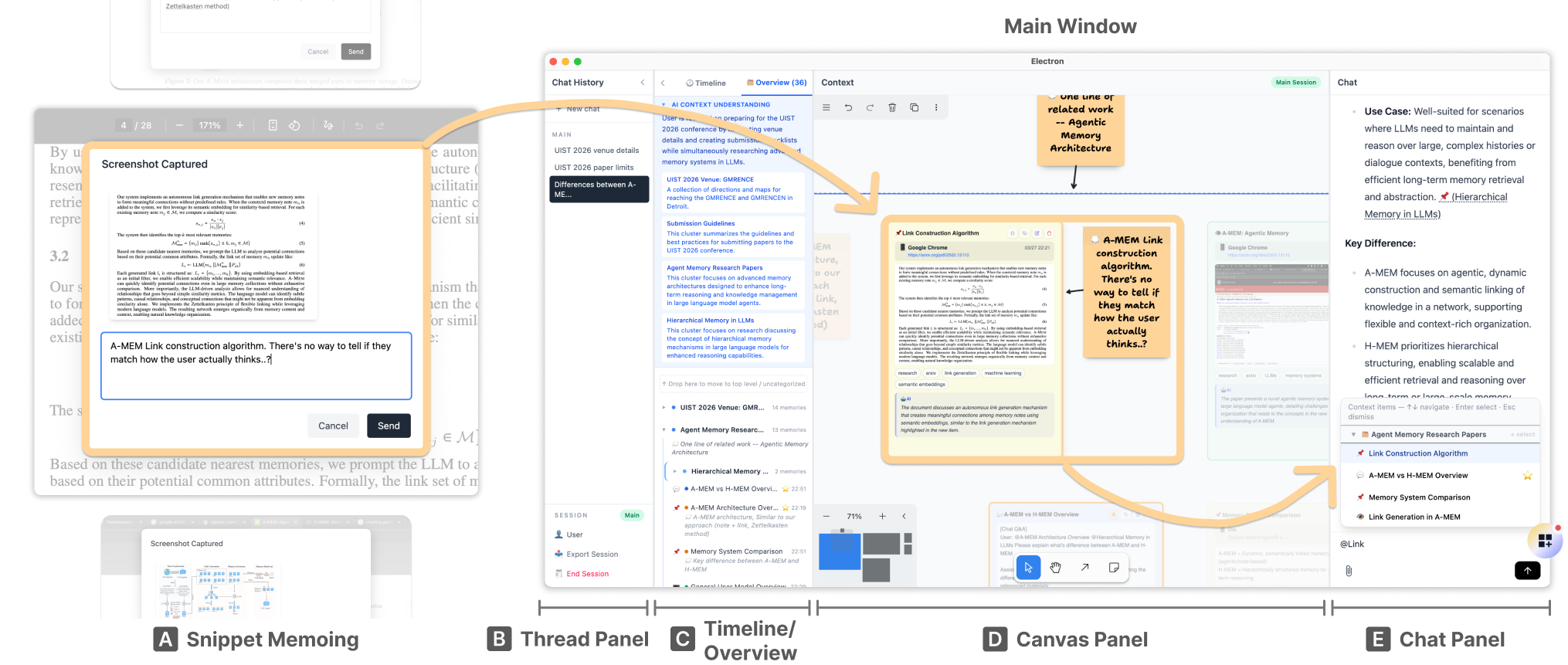}
  \caption{Overview of \sysname{}. Users capture their in-situ thoughts during their task via Snippet Memoing (A), which are fed into the AI's context. This context is visualized on the Canvas Panel (D) in an inspectable, correctable form, allowing users to review, reorganize, and refine how the AI has represented their task context. Users can also navigate this context through a Timeline or hierarchical Overview (C), and interact with an AI agent grounded in this context via the Chat Panel (E).}
  \Description{Teaser figure showing the full workflow of \sysname{} with five labeled components. (A) Snippet Memoing shows a floating pop-up over a research paper where the user has captured a screenshot and written an in-situ thought about a link construction algorithm. Below it, a second snippet capture is partially visible. The Main Window contains four panels: (B) Thread Panel on the far left lists conversation threads. (C) Timeline/Overview in the center-left shows captured memories chronologically with tabs for Timeline and Overview views. (D) Canvas Panel in the center displays memory cards organized into labeled groups with connecting annotations and sticky notes, showing the AI's structured representation of the user's task context in an inspectable, correctable form. (E) Chat Panel on the right shows a conversation where the user asks about agentic memory architectures, and the assistant responds with a detailed comparison grounded in the captured context, referencing specific memory items.}
  \label{fig:teaser}
\end{teaserfigure}

\begin{abstract} 
During complex knowledge work, people engage in iterative sensemaking: interpreting information, connecting ideas, and refining their understanding. Yet in current human–AI collaboration, these cognitive processes are difficult to share and organize for AI. They arise in situ and are rarely captured without interrupting the task, and even when expressed, remain scattered or reduced to system-generated summaries that fail to reflect users' cognitive processes. We address this challenge by enabling AI context that is grounded in users' cognitive traces and can be directly inspected and revised by the user.
We first explore this through a probe system that supports in-situ snippet memoing, allowing users to easily share their cognitive moves. Our study (N=10) highlights the value of capturing such context and the challenge of organizing it once accumulated. We then present \sysname{}, which supports users in inspecting and refining these contexts to better reflect their understanding of the task. Our evaluation (N=12) showed that \sysname{} improved task awareness, thought structuring, and users' sense of authorship and control, with participants preferring snippet-grounded AI responses over non-grounded ones (78.1\%). We discuss how capturing and organizing users' cognitive context enables AI as a context-aware collaborator while preserving user agency.
\end{abstract}
\maketitle
\section{Introduction}
Much of the reasoning that shapes complex tasks emerges in fragmented, moment-to-moment cognitive moves~\cite{schon1992reflective}. As users read a paper, compare alternatives, or piece together information across sources, they continuously form and revise judgments---why a finding feels significant, how options are being weighed, or what criteria have just shifted. However, these fleeting thoughts are difficult to share and organize into current AI systems. They arise during the task, but articulating them requires interrupting the task to write a prompt. Even when expressed, they remain scattered across chat turns rather than organized into a structured, user-accessible form. As a result, users struggle to reconstruct what has been considered or why, while the AI continues to operate on context that is misaligned with the user’s current understanding of the task, offering suggestions that are plausible but no longer useful.

While recent work shows that passive observation of users' on-screen activity can provide rich contextual signals~\cite{wang2025openhandsopenplatformai, shaikh2025GUM}, behavioral traces reveal what a user did and where they spent time, but cannot capture the user's internal criteria or interpretations behind those actions---the \textit{why} and \textit{how} of what they did~\cite{yang26guide}. 
Similarly, recent LLM systems' automatic memory or context summarization~\cite{chatgptmemory, claudecode} distill interaction history into user-readable summaries. However, these summaries reflect system-generated abstractions rather than how users prioritize information or organize relationships between ideas, making it difficult for users to inspect or understand how their task is currently represented.

We frame this as two interrelated challenges. First, a \textbf{capturing problem}: users lack lightweight ways to externalize their in-situ reasoning as they work, so fleeting judgments, comparisons, and interpretations disappear rather than becoming part of the shared context. Second, a \textbf{visibility problem}: even when some context is shared, it remains buried in chat history or reduced to system-generated summaries, leaving neither side with a clear picture of how the user's thinking about the task has progressed. These challenges echo long-standing insights from HCI and cognitive science: complex work is shaped by in-the-moment reasoning that disappears if not externalized~\cite{schon1992reflective}, and understanding only deepens when captured thoughts can be organized and revisited~\cite{Hollan2000Distributed, zhang1997nature, Norman1991CognitiveArtifacts}.

Building on these insights, we propose a different approach to context sharing in human–AI collaboration, giving users lightweight affordances to externalize their reasoning in-situ and to inspect and refine the AI context.
We first explored this idea through a probe system that allowed users to capture screen snippets and attach short in-situ memos directly into an LLM's context.
We deployed the probe with 10 experienced LLM users on real-world sensemaking tasks and found that in-situ snippet memoing enabled richer, lighter, and more forward-looking context sharing than prompting. However, the study also revealed that capturing alone was insufficient.
As snippets accumulated, participants wanted to organize these evolving cognitive traces---grouping related ideas, tracking changing criteria, and connecting information across sources---while also understanding how this evolving context was represented and used by the AI.

To address these challenges, we present \sysname{}, a system that externalizes AI context as a user-inspectable and correctable representation grounded in users’ cognitive processes. 
Building on the probe, \sysname{} introduces a canvas-based context space that organizes AI context constructed from users' in-situ snippet memos and passive computer-use observations.
The canvas serves both as an inspectable view of the context AI uses to ground its responses and as the user's sensemaking workspace. Because this context is continuously assembled from users' ongoing task activity and in-situ thoughts, inspecting and refining AI context becomes a natural part of doing the task.

We evaluated \sysname{} in a within-subjects study with 12 participants on real-world exploratory tasks. Our results show that \sysname{} significantly improves users' task awareness, thought structuring, perceived AI understanding, and their sense of authorship and control compared to a baseline condition. 
Participants also consistently preferred AI responses generated with snippet context over those without (78.1\%), demonstrating that users' in-situ thoughts provide meaningful grounding for more relevant AI support.
These findings point to a broader opportunity: when AI context is grounded in user-authored cognitive traces and made inspectable and correctable, human–AI collaboration can become more transparent and more aligned with users' ongoing cognitive processes during the task.
At the same time, our study surfaces important trade-offs between the effort of externalization, perceived usefulness, and users' sense of agency. 
Based on these findings, we discuss how future systems can capture user context with minimal burden while supporting its organization, inspection, and reuse over long-term human–AI collaboration.

\section{Related Work}

\subsection{Context Sharing for Human-AI Collaboration}
Effective human-AI collaboration relies on shared context between users and AI systems. To provide relevant and aligned support, AI systems must understand not only users’ explicit requests but also the broader context and underlying intent that shape those requests.
Prior work has explored various approaches to how user context can be constructed and utilized in human-AI collaboration. 
One line of work focuses on interaction techniques to better elicit and align user intent, including scaffolding prompt formulation~\cite{brade2023promptify}, supporting iterative refinement~\cite{Wang2024PromptCharm}, and representing intent as structured or manipulable units and as evolving, revisable configurations across interaction~\cite{gmeiner2025intenttagging, kim2025intentflow}.
These approaches improve how users articulate and maintain their intents, but still rely primarily on explicitly expressed inputs.
Another line of work focuses on how AI context is constructed, maintained, and utilized.
In context engineering, researchers have explored incorporating broader context, such as retrieved documents, external knowledge, and interaction history, into model inputs to improve grounding and response quality~\cite{mei2025survey}. In parallel, recent work has enabled users to supply and curate contextual knowledge that can be dynamically integrated during interaction~\cite{zhao2025knoll}, or embedding AI into shared workspaces to make interactions more visible and situated~\cite{lee2025choir}. 
Beyond supplying context, some work makes the AI's grounding context itself explicit and user-manageable by reifying past conversations into memory objects that users can retrieve and reuse~\cite{yen2024memolet}, or letting users commit new intent into a specification that grounds the AI's responses, with support for resolving conflicts that arise~\cite{vaithilingam2025semantic}.
Agent-based approaches further incorporate continuous interaction traces and environmental signals, enabling AI systems to proactively gather and utilize context beyond explicit user input \cite{yang2025contextagent, lu2024proactive}.
Our work builds on this direction but focuses on a different form of context. Rather than retrieved information, prior conversations, user-authored specifications, or behavior traces, we investigate how AI context can be continuously assembled from users' ongoing task activity and cognitive processes. We further explore how this context can be represented in a form that users can inspect, organize, and refine as they work, supporting both users' evolving understanding of the task and the AI's grounding.

\subsection{Externalizing In-situ Cognitive Moves in Knowledge Work}
Externalizing thoughts plays a central role in knowledge work, enabling people to offload cognitive effort and refine their understanding over time \cite{Hollan2000Distributed, Norman1991CognitiveArtifacts}. Prior work further suggests that much of this reasoning unfolds as in-the-moment activity (e.g., reflection-in-action), where users continuously interpret, compare, and adjust their understanding during a task \cite{schon1992reflective}.

Prior systems have explored supporting thought externalization through different interaction designs. Early approaches drew on think-aloud methods that externalize ongoing thought processes~\cite{krosnick2021think}. Building on this idea, recent systems support in situ captures integrated with ongoing activity. For example, \textit{PointAloud}~\cite{gmeiner2026pointaloud} enables users to articulate reasoning alongside pointer-based interactions, while \textit{ClearFairy}~\cite{son2025clearfairy} captures intermediate reasoning through in-situ questioning and rationale inference. 
\textit{LiquidText}~\cite{tashman2011liquidtext} and related annotation-based work~\cite{marshall1997annotation} support linking ideas within reading workflows; and \textit{Fuse}~\cite{kuznetsov2022fuse} enables collecting and comparing information during browser-based exploration. More recently, \textit{Orality}~\cite{li2026orality} further reduces the effort of capturing early-stage thoughts by transforming them into structured representations.
Despite these advances, prior approaches do not support using in-situ cognitive moves as a shared resource for working with AI, limiting their role in maintaining aligned task context over time; we address this by enabling their capture, organization, and collaborative use in human–AI interaction.

\subsection{Organizing Externalized Thoughts in Knowledge Work}
Externalizing thoughts alone is insufficient unless they can be organized and revisited over time. Schön’s notion of \textit{reflection-on-action} emphasizes the importance of returning to externalized artifacts to interpret, restructure, and deepen understanding \cite{schon1992reflective}. Similarly, prior work on external representations shows that organizing information into structured forms reduces cognitive load and makes implicit relationships more explicit \cite{zhang1997nature, ahrens2022take}.

Prior work has explored supporting such organizations in various ways. Systems such as \textit{ConceptScope} and \textit{ConceptEVA} support iterative refinement of users’ conceptual understanding, helping users evolve their interpretation of a topic over time \cite{zhang2021conceptscope, zhang2023concepteva}.
In the context of LLM interaction, systems such as \textit{Sensecape} and \textit{Graphologue} structure outputs into visual or relational representations for comparison \cite{suh2023sensecape, jiang2023graphologue}, while \textit{Luminate} and \textit{Policy Maps} map outputs into structured spaces to support systematic analysis \cite{suh2024luminate, lam2025policy}. Beyond LLM outputs, \textit{Orca} treats webpages as malleable materials that users and AI can collaboratively organize into a dynamic browser-level workspace \cite{jiang2025orca}. 
In scholarly sensemaking, \textit{ScholarMate} and \textit{Synergi} integrate AI into visual workspaces for organizing literature and qualitative codes to support collaborative analysis and synthesis~\cite{ye2025ScholarMate, kang2023Synergi}.
Recent work also explores shared representations for organizing evolving context in collaboration~\cite{zhang2025ladica}.
While these systems support organizing information, they typically focus on externally available content such as model outputs, documents, or literature.
Our work instead focuses on organizing users' own evolving task context by incorporating users' in-situ thoughts and ongoing task activity into a shared context, grounding AI support in users' cognitive processes.
\section{Design Exploration: Understanding Cognitive Externalization via Snippet Memoing}\label{section:design_explore}
We begin by exploring how users share their in-situ thoughts with an AI as they work. Much of this reasoning---interpreting information, comparing options, and forming tentative judgments---emerges in the moment, but these fleeting cognitive moves are rarely captured or shared with the AI. In particular, prompting requires users to reformulate these thoughts as explicit instructions, which introduces friction and often filters out process-level reasoning. To explore this, we introduce \textbf{\textit{snippet memoing}}, a lightweight interaction method that allows users to capture and share fragments of information on screen along with their associated thoughts in situ, directly into the LLM's context.

\subsection{Research Probe}\label{section:probe}
\begin{figure*}
    \centering
    \includegraphics[width=1\linewidth]{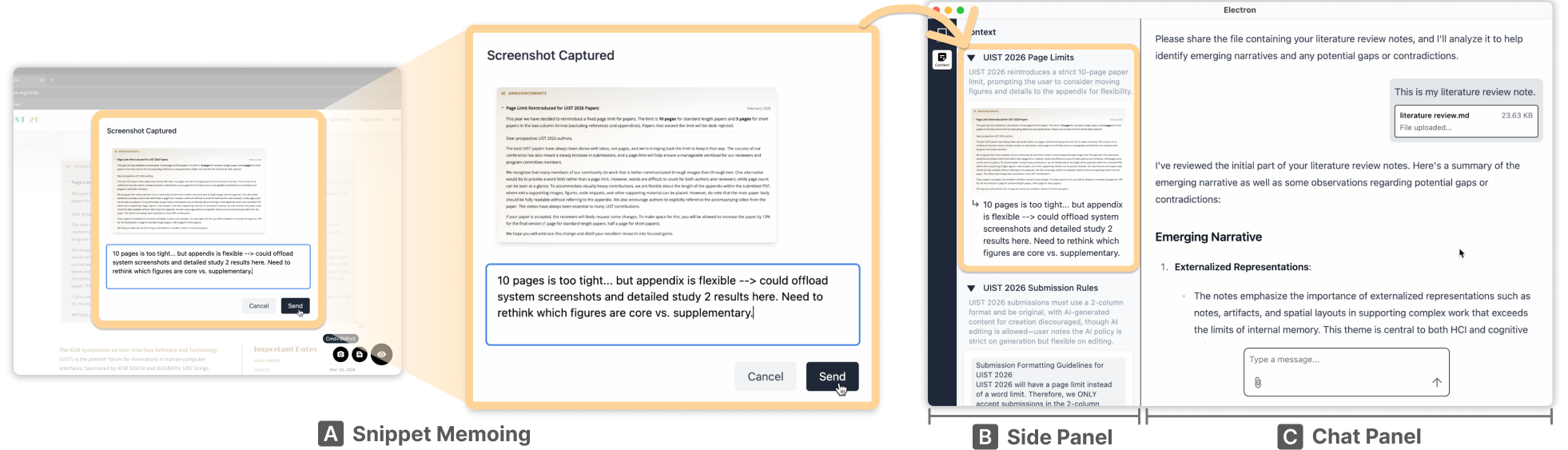}
    \caption{The probe system for snippet memoing. (A) When a user captures a screenshot snippet, a pop-up appears showing the captured content alongside a text field for attaching an in-situ memo. (B) Sent snippet–memo pairs accumulate in a side panel, where each entry displays the captured content and the user's memo, alongside an AI-generated title and summary reflecting how the system interpreted the snippet. (C) The chat panel on the right allows users to interact with the LLM, which generates responses grounded in the accumulated context.}
    \Description{Annotated screenshot of the probe system for snippet memoing with three labeled components. (A) Snippet Memoing shows a pop-up overlay displaying a captured screenshot with a text field below where the user has typed an in-situ memo about appendix structure. Cancel and Send buttons appear at the bottom. (B) Side Panel shows a chronological list of sent snippet--memo pairs, each with an AI-generated title and summary reflecting the system's interpretation of the captured content. (C) Chat Panel on the right shows a conversation where the user asks the LLM to analyze their literature review notes, and the assistant responds with a structured summary organized under headings, grounded in the accumulated snippets.}
    \label{fig:probe}
\end{figure*}
We developed a desktop-based LLM application as a probe that enables users to capture and share their in-situ thoughts while working on a task. The application registers global keyboard shortcuts (\texttt{Cmd/Ctrl+Shift+S} for screenshots, \texttt{Cmd/Ctrl+Shift+C} for text) that can be triggered from any application, without switching applications or interrupting users' workflow. A pop-up immediately appears showing the captured content alongside a text field where users can attach a short memo: a brief thought, interpretation, or rationale tied to what is on screen at that moment (Figure~\ref{fig:probe}A). 

Once sent, these snippet–memo pairs accumulate in a persistent side panel, where each entry displays the captured content and the user's memo, alongside an AI-generated title and summary reflecting how the system interpreted the snippet (Figure~\ref{fig:probe}B). All accumulated snippets are included in the LLM's context window, and users can interact with the LLM through a chat panel to ask questions or request assistance grounded in this context (Figure~\ref{fig:probe}C). The system was built with Electron~\cite{electron} and TypeScript, compatible with macOS and Windows, and used \texttt{gpt-4o-2024-08-06} as the underlying LLM. Participants installed the system on their personal laptops to preserve their natural work environment.

\subsection{Participants and Procedure}
We recruited 10 participants (5 male, 5 female; age $\mathit{M}=26$, age $\mathit{SD} = 1.87$) with extensive daily LLM experience (e.g., ChatGPT, Claude, Gemini), with four participants using them every day and six using them 2-5 times a week. Participants were recruited through online recruitment postings on our university community platform. 
The task was designed to elicit exploratory, multi-source sensemaking, where users interpret information, compare options, and iteratively refine their understanding, allowing in-situ thought capture to naturally arise. Participants explored potential labs and advisors to build a shortlist, and then drafted a cold email to one selected professor.
To ensure ecological validity, we pre-screened participants who were actively considering applying to MS, Ph.D., or postdoctoral programs, confirming that the task would be personally relevant and motivating.
Each session began with a 10-minute system tutorial, followed by a 30-minute task session in which participants used the system to complete the task while we observed their process, and concluded with a 20-minute post-survey and semi-structured interview. The survey included NASA-TLX and 7-point Likert-scale items on overall satisfaction, perceived context utilization, and burden of sharing the context. The interview explored how this form of interaction differed from conventional prompting, what types of thoughts participants captured and why, and what design opportunities or pain points they perceived. All sessions were screen-recorded and transcribed for later analysis. Participants were compensated KRW 20,000 ($\approx$ USD 14) for their time, and the study was approved by our institutional IRB.

\subsection{Findings: How Snippet Memoing Supports Cognitive Externalization}
Our findings show that snippet memoing enables different context sharing compared to traditional prompting, capturing not only what users explicitly ask, but also their ongoing interpretations, partial conclusions, and task-relevant reasoning as they unfold.

\subsubsection{\textbf{Expressing user-framed perspectives and forward-looking thoughts}}
Snippet memoing enabled participants to externalize not just \textit{what} they were looking at, but \textit{how} they were making sense of it in the moment. By attaching a memo to each captured snippet, users expressed their own interpretations, emerging criteria, and tentative judgments---what felt significant, what remained uncertain, or how it might be used. This created a form of user-framed context, where each snippet reflected the user's cognitive process at that moment. P2, P5, and P8 emphasized that this allowed the system to better align its responses with their intentions. This benefit was also reflected in the post-survey: participants rated the item ``I felt that the system made good use of the information I provided with a better understanding of my context'' with an average of 6.0/7 (SD = 0.67).
Participants also used snippets to capture forward-looking thoughts as they arose. Rather than waiting to formulate a concrete question, they externalized observations and partial ideas that they anticipated to be relevant later, even when they were not yet sure how (P4–6, P8). 
This distinction between ``what I want to remember or use later'' and ``what I want to ask'' gave participants greater flexibility in their interactions, in contrast to the prompt-only method, where context must be compressed into a single turn.

\subsubsection{\textbf{Lower barrier to externalizing thoughts in situ}}
Participants (P2–3, P5–6, P8–10) emphasized that snippet memoing lowered the cognitive and interactional cost of sharing context. Rather than pausing to compose elaborate prompts or switch applications, they could capture and annotate their thoughts continuously while working. P6 and P8 noted that because they could explicitly choose where and how
to share their thoughts, there was less psychological burden compared to composing a full prompt. This ease was also reflected in the post-survey: when asked \textit{``The way I shared what I was looking at or thinking about to the system didn't feel burdensome''}, participants rated it with an average of 6.6/7 (\(SD = 0.66\)). NASA-TLX scores further confirmed that the interaction did not impose substantial workload: mental demand ($M=2.5$), effort ($M=2.4$), and frustration ($M=1.3$) all remained low, while perceived performance was high ($M=5.6$).

\subsubsection{\textbf{Captured snippets as cognitive scaffolds}}
Finally, participants also appreciated that these captured snippets served as cognitive scaffolds for their own task performance. P2-4, P6, and P10 reported that being able to look back on what they had seen and thought about helped them better navigate the task, maintain continuity, and build on earlier insights as the task progressed. Rather than merely supporting the system, capturing and revisiting accumulated snippets made users' own reasoning more visible to themselves throughout the task.
\begin{figure*}
    \centering
    \includegraphics[width=1\linewidth]{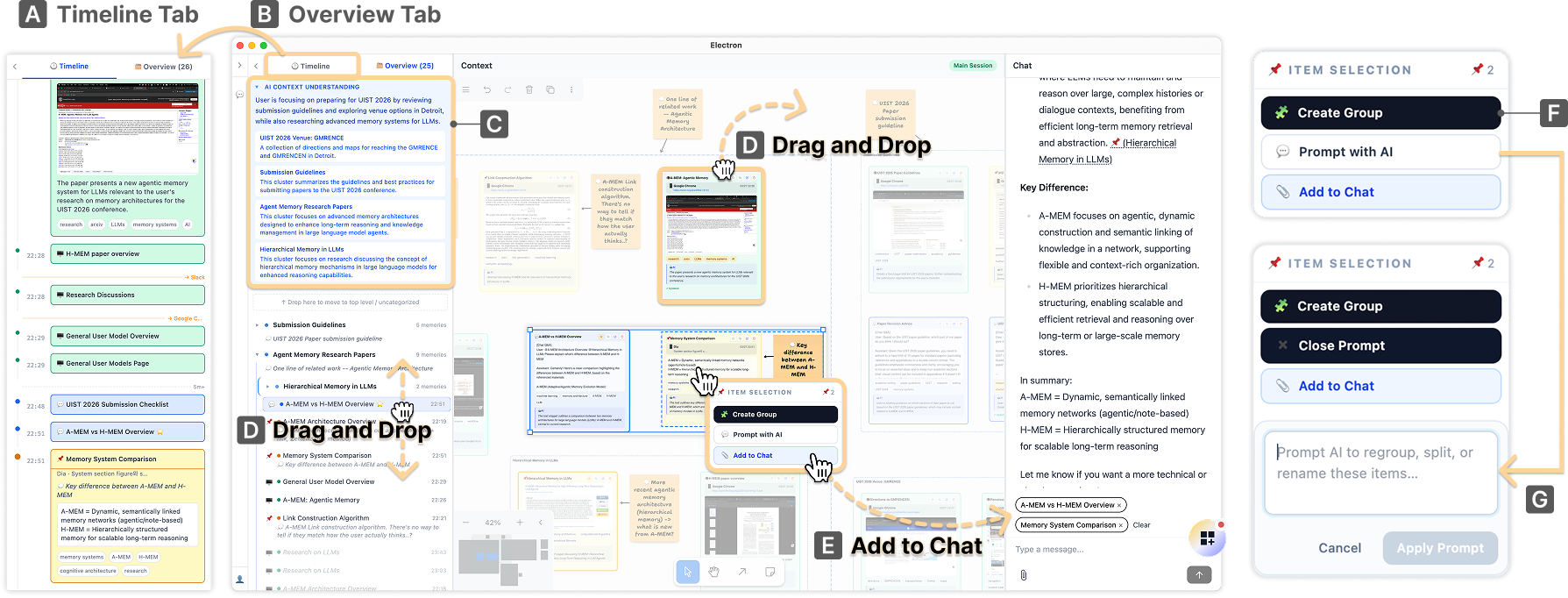}
    \caption{Overview of the main window of \sysname{}. The memory overview provides both timeline (A) and hierarchical overview (B) views of captured context. The system maintains an AI-generated context summary (C) of the user’s current focus. Users can reorganize items and restructure groups via drag-and-drop on the canvas (D), and by right-clicking selected items to group them into new branches or invoke AI-assisted grouping (F, G). Selected items can also be added to chat (E).}
    \Description{Annotated screenshot of the main window of \sysname{}, showing seven labeled components. (A) Timeline Tab displays captured memories chronologically with color-coded source indicators. (B) Overview Tab shows a hierarchical tree view of branches and their member memories. (C) AI-generated context summary provides a concise description of the user's current focus. The central canvas displays memory cards organized into groups, with (D) drag-and-drop interactions shown for reorganizing items between groups. (E) Add to Chat demonstrates attaching a selected memory to the chat conversation via right-click. (F) Right-click context menu on selected items offers Create Group, Prompt with AI, and Add to Chat options. (G) AI prompt interface allows users to provide natural language instructions for regrouping selected items.}
    \vspace{-5pt}
    \label{fig:system}
\end{figure*}
\subsection{Design Implications: User-Inspectable and Correctable AI Context}
While snippet memoing enabled richer and more user-framed context sharing, the study also surfaced a key limitation. As snippets accumulated, participants found it difficult to keep track of what they had captured and how it was reflected in the AI’s understanding. This made it harder to revisit prior thoughts, see what had already been considered, and ensure that the AI remained aligned with what they actually found important (P2–P10). Importantly, participants did not simply express a need for better organization of snippets. Rather, they wanted the AI’s context itself to reflect their thinking in a way that they could see and influence. At the same time, relying solely on users to manage accumulated context made it difficult to maintain alignment with the AI. Participants instead preferred system support while retaining control over how their context is represented (P6–P8).
Based on these observations, we derive two key design implications for supporting the structuring and use of captured context in human–AI collaboration:
\textbf{DI1. Ground AI context in users’ cognitive processes.}
Systems should incorporate users’ in-situ thoughts as first-class inputs in forming AI context, so that the resulting context reflects what users consider important and how they interpret information. Rather than relying solely on system-inferred signals, this allows the AI’s understanding to be directly shaped by the user’s perspective.
\textbf{DI2. Make AI context inspectable and correctable by users.}
The AI’s context should be externalized in a form that users can directly inspect and revise. This allows users to verify how their inputs are being interpreted, identify mismatches, and adjust the context when needed~\cite{shneiderman1983direct}. By keeping users in the loop, systems can maintain alignment with users’ thinking while preserving their sense of authorship and control.
\section{\sysname{}} \label{section:contexty}
Based on the design implications, we developed \sysname{}, a system that externalizes AI context as a user-inspectable and correctable representation (DI2) grounded in users’ cognitive processes (DI1). It extends snippet memoing (in Sec \ref{section:probe}) by incorporating users’ in-situ thoughts as first-class inputs to AI context, and surfaces this context through a canvas-based representation that makes it inspectable and correctable by users. 
The canvas serves both as an inspectable view of the context grounding AI responses and as a workspace where users organize and refine their evolving understanding of the task. Because the context is continuously assembled from users' ongoing task activity and in-situ thoughts, refining the canvas simultaneously refines the AI's grounding.
Using \sysname{}, as users work across different sources, they can capture snippets along with their in-situ thoughts (Figure~\ref{fig:teaser}A); \sysname{} integrates these into the AI’s context and surfaces them on the canvas (Figure~\ref{fig:teaser}C, D). Users can inspect and refine this canvas at any time, and converse with an AI assistant (Figure~\ref{fig:teaser}E) grounded in this context throughout their workflow.

\subsection{Context Capture}
\begin{figure}
    \centering
    \includegraphics[trim=0 3pt 0 3pt, clip, width=1\linewidth]{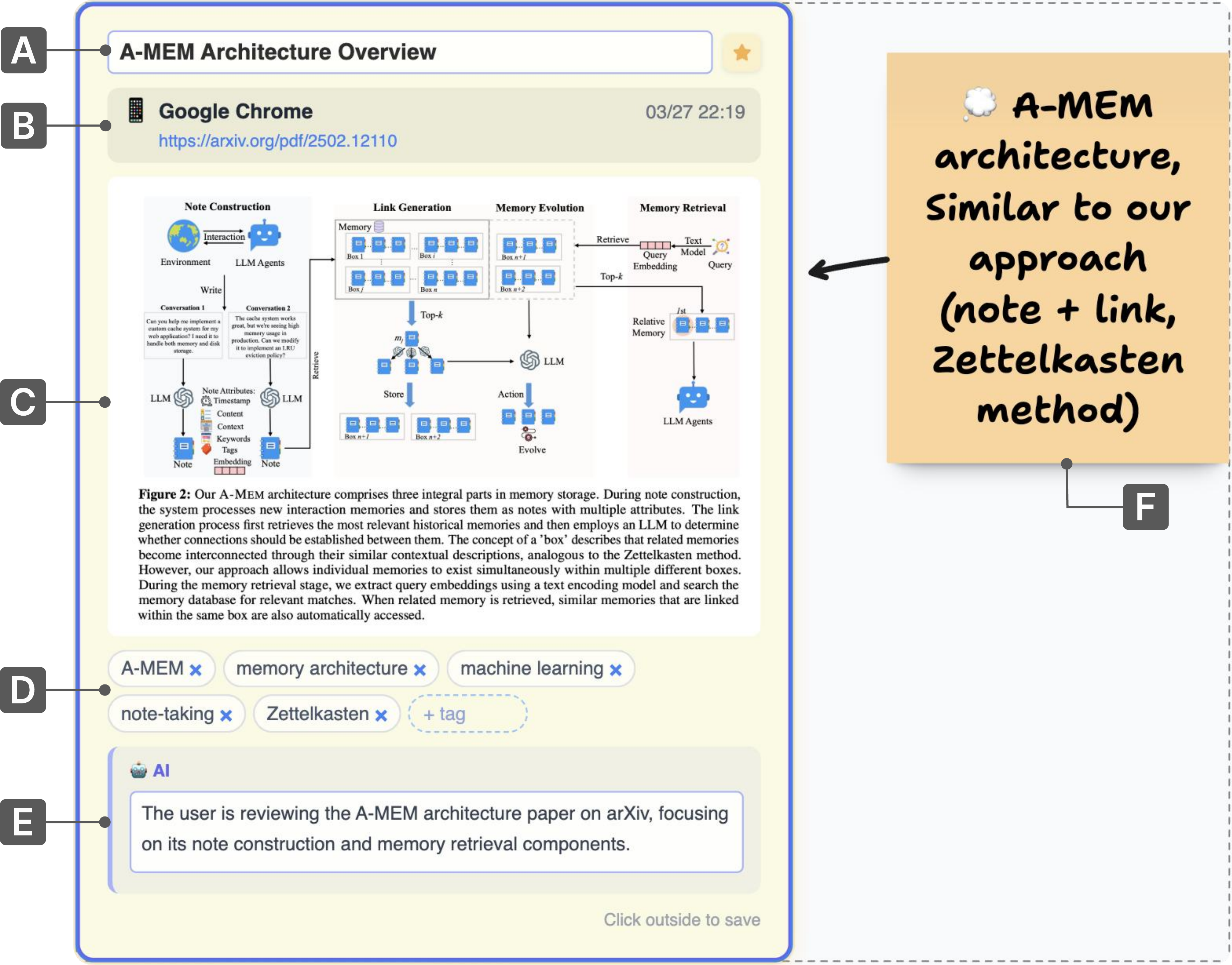}
    \caption{A memory card on the \sysname{} canvas (edit view). Each card displays (A) an AI-generated title, (B) provenance information including the source application, timestamp, and URL, (C) the raw captured content, (D) AI-generated tags, and (E) an AI-generated user's context reflecting how the system interprets the user's current activity at the time of capture. For snippet captures, (F) the user's in-situ memo is also displayed as a sticky note alongside the card.}
    \vspace{-10pt}
    \Description{A memory card displayed on a canvas with labeled components. (A) At the top, an AI-generated title is shown. (B) Below it, provenance information includes the source application and URL. (C) The center of the card displays the main captured content as a document preview. (D) Beneath the content, a set of AI-generated tags is presented. (E) At the bottom, a text box shows an AI-generated description of the user’s current context. (F) To the side of the card, a sticky note displays the user’s in-situ memo associated with the captured content.}
    \label{fig:memory-card}
\end{figure}
\sysname{} captures user context through three complementary channels---\textit{snippet memoing}, \textit{computer-use observations}, and \textit{chat}.
Building on prior work showing that passive observation of users' on-screen activity can provide rich contextual signals for both LLMs and computer-use agents~\cite{shaikh2025GUM, lu2024proactive, shaikh2026learningactionpredictorshumancomputer, lam2025just}, \sysname{} supports \textbf{computer-use observation}. Observations provide broad situational awareness of users’ ongoing activities without requiring additional effort, enabling the system to capture background task context across applications and over time. \sysname{} monitors global input events and automatically captures a full-screen screenshot whenever the user clicks or presses \texttt{Enter} key outside \sysname{}.
Users can disable passive observation at any time if preferred.
\textbf{Snippet memoing} follows the same interaction design as the research probe (Section~\ref{section:probe}, Figure~\ref{fig:probe}A). Finally, \textbf{chat} exchanges between the user and LLM assistant are also treated as a first-class context channel.
Each user prompt and its corresponding AI response are stored alongside snippets and observations as part of the shared task context, allowing future responses to build on the user's prior interactions.

\subsubsection{\textbf{Processing and Display}} 
For snippets and observations, the system extracts provenance metadata (\texttt{appName}, \texttt{windowTitle}, \texttt{url}) at capture time using OS-level APIs. Each captured item, along with its metadata, is analyzed by an LLM and rendered as a card on the canvas (Figure~\ref{fig:memory-card}), displaying (A) an AI-generated title, (B) provenance information including the source \texttt{appName}, \texttt{url}, and timestamp, (C) the raw captured content, (D) AI-generated tags, and (E) an AI-generated user’s context reflecting how the system interprets the user's current activity and intent. For snippet captures, the user's in-situ memo is additionally shown alongside the card (F). Cards are color-coded by source type: yellow for snippets, green for observations, and blue for chat. Users can edit any card via an edit button to correct or refine its content.
Near-identical observation screenshots are deduplicated via perceptual hashing, and redundant observations are automatically hidden on the canvas based on semantic similarity to existing visible memories (see Appendix~\ref{appendix:observation-filtering} for details).

\subsection{Context Memory Structure}
As captured items from \textbf{observation}, \textbf{snippets}, and \textbf{chat} accumulate over time, \sysname{} organizes them into a memory structure to effectively incorporate them into AI's context. Inspired by recent agentic memory architectures, which organize LLM memories into Zettelkasten-inspired note-link structure or hierarchical structure~\cite{xu2025Amem, sun2026h}, \sysname{} organizes captured items into a hierarchical memory structure $T = (M, B, L)$, where $M$ is a set of \textit{memories} (individual captured items), $B$ is a set of \textit{branches} (semantic groups of memories with LLM-generated names and summaries), and $L$ is a set of \textit{links} $l = (m_i, m_j, \sigma)$ capturing relationships across memories, each weighted by a relevance score $\sigma \in [0,1]$.
Each memory $m \in M$ is assigned to a branch $b \in B$ (or left unassigned).
Each link $l \in L$ is inferred as new memories arrive and is used internally to organize the memory structure.

\subsubsection{\textbf{Automatic Placement}} 
As new memories are added, the system automatically places each item into the existing structure to maintain a coherent organization of the AI context without requiring manual effort. When a new memory $m$ is added, the system retrieves the top-$k$ semantically related existing items using an LLM, along with relevance scores $\sigma_i \in [0,1]$, which constitute the links in $L$.
The scores are aggregated at the branch level: $S(b) = \sum_{m_i \in b} \sigma_i$.
The memory is assigned to $b^* = \arg\max_{b \in B} S(b)$ if $S(b^*) > 0$; otherwise, a new branch is created from the memory together with its related unassigned items, or the memory is left unassigned if there are none.

\subsubsection{\textbf{Memory Evolution}} 
As new memories are added, the system continuously refines the existing memory structure to keep it aligned with the user's task context. When a new memory provides additional context, the LLM suggests metadata updates for related existing items, including refined tags and context descriptions. When related memories are updated, an ``Updated'' badge is shown on the affected cards, signaling to the user that the system has revised its interpretation of previously captured content.

\subsubsection{\textbf{Users' Reorganization}}
Users can directly modify the memory structure $T$ through multiple interaction methods on the canvas. They can drag-and-drop items (Figure~\ref{fig:system}D) to move memories ($m$) across branches ($b$). Users can also select a set of items and, via right-click, group them into a new branch (Figure~\ref{fig:system}F). Alternatively, users can provide natural language instructions (e.g., ``group these by project'') to let the system propose and apply structural updates (Figure~\ref{fig:system}G).

\subsection{Context-Augmented Chat}
\begin{figure}
    \centering
    \includegraphics[trim=0 3pt 0 3pt, clip, width=1\linewidth]{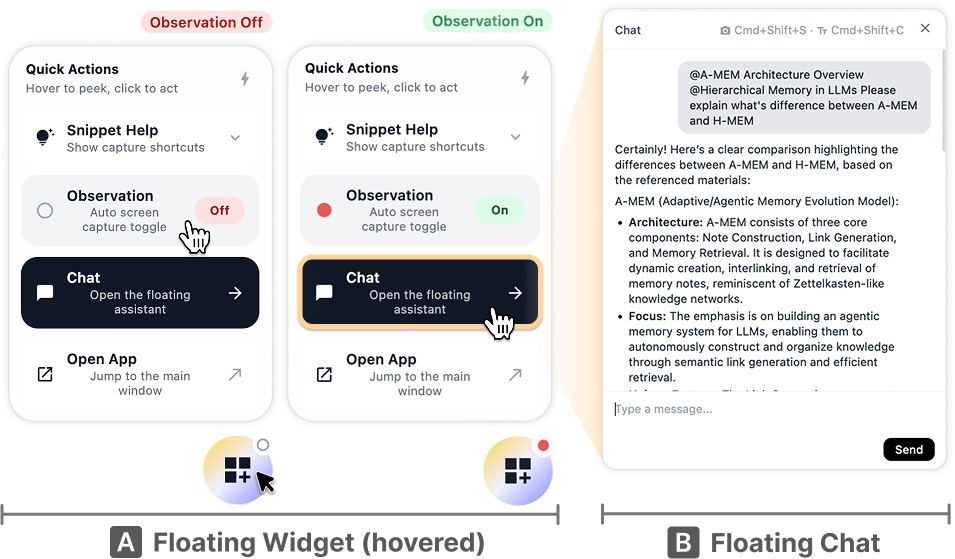}
    \caption{Floating widget (A) provides quick access to snippet memoing, observation toggle, and (B) a mirrored chat.}
    \vspace{-10pt}
    \Description{A floating widget (A) with quick actions for snippet memoing, observation on/off toggle, and access to a mirrored chat panel. Next to it, a floating chat window (B) mirrors the main system chat, showing a conversation with an AI assistant. Both remain accessible over other applications, allowing users to capture context and interact with the AI without returning to the main system interface.}
    \label{fig:floating}
\end{figure}
\sysname{} provides a chat interface that grounds AI responses in the user's accumulated task context. The interface is available both as a panel fixed alongside the canvas and as a floating widget (Figure~\ref{fig:floating}B) that can be positioned anywhere on the screen, with both views remaining synchronized.

Users can explicitly bring specific memories into a conversation by @-mentioning them or by right-clicking a card on the canvas and selecting ``Add to Chat'' (Figure~\ref{fig:system}E). Beyond explicit references, the system automatically retrieves memories relevant to the user's query, taking into account lexical overlap, tag matches, and recency (see Appendix~\ref{appendix:retrieval} for details). The retrieved memories are prepended to the user's message with source metadata, allowing the AI to respond with awareness of the user's broader task context.

\sysname{} also maintains an \textbf{AI Context Understanding}, a concise LLM-generated summary of the user's current activities and focus (Figure~\ref{fig:system}C), derived from the memory structure by analyzing group structures, recent memories weighted by source type (with snippets weighted more heavily than observations), and strong links between memories, which may span branches and indicate concurrent work across domains.

\subsection{User Interface}
\sysname{} presents a four-panel layout (Figure~\ref{fig:teaser}): (1) a \textbf{thread panel} for managing multiple conversation threads and session lifecycle, (2) a \textbf{memory overview panel} displaying captured memories either chronologically as a timeline view or as a hierarchical overview of the context memory structure, switchable via tabs, (3) a \textbf{canvas panel} providing an interactive spatial visualization of the context memory structure, and (4) a \textbf{chat panel} for conversational interaction with the LLM assistant. In addition, a \textbf{persistent floating widget} remains visible above all applications regardless of the user's current task. It provides quick access to snippet memoing shortcuts, observation toggling, and a mirrored chat panel so users can interact with the AI without switching back to the system.
The canvas renders each branch as a group of memory cards, with child branches displayed as nested groups within their parent---allowing hierarchical task structure to be reflected spatially. Users can reorganize the structure through drag-and-drop on either the canvas or the hierarchical overview panel, moving individual memories or entire groups to reassign their place in the hierarchy.

\subsection{Implementation}
\sysname{} is implemented as a cross-platform desktop application using Electron, TypeScript, and React. On the backend, core services handle context capture, memory management, and chat orchestration, with session data persisted locally. On the frontend, the interactive canvas is built on tldraw~\cite{tldraw}, and the floating widget is rendered as an always-on-top overlay window. LLMs are used throughout the system pipeline, all powered by \texttt{gpt-4o-mini-2024-07-18}: content analysis, memory placement, related-item retrieval, and AI-assisted reorganization, except for the main conversational chat, which uses \texttt{gpt-4.1-2025-04-14}, orchestrated via OpenAI's Assistants API with persistent threads. All prompts used for LLM interactions are provided in the supplementary material.

\begin{figure*} [t]
    \centering
    \includegraphics[width=1\linewidth]{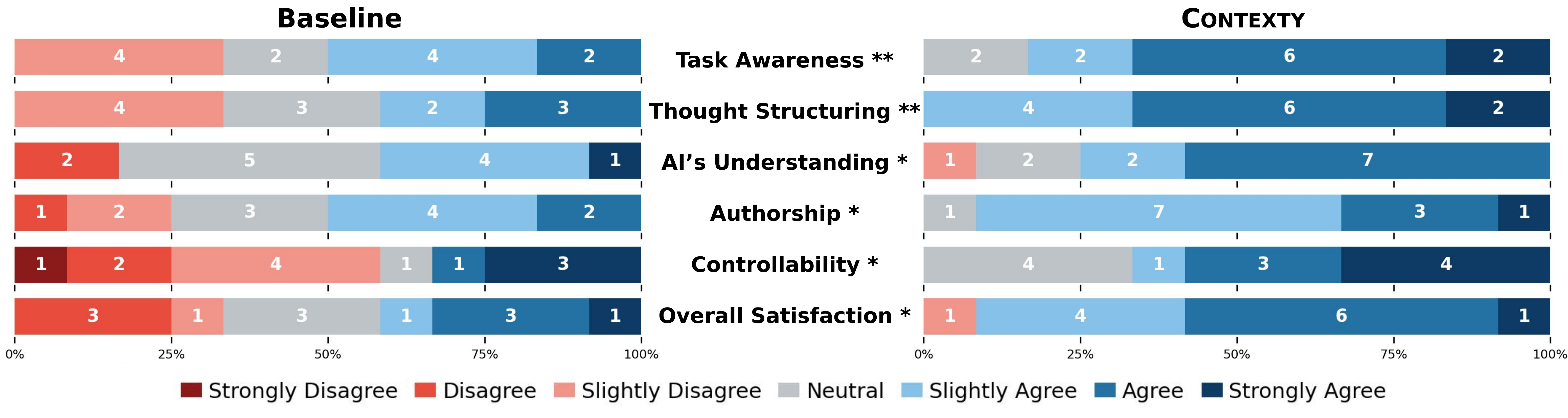}
    \caption{Post-condition survey results comparing \sysname{} and the baseline across six dimensions (7-point Likert scale). Asterisks denote statistical significance from Wilcoxon signed-rank tests (*$p<.05$, **$p<.01$)}
    \Description{Diverging stacked bar charts comparing post-condition survey responses between Baseline (left) and \sysname{} (right) across six dimensions on a 7-point Likert scale. \sysname{} receives substantially more Agree and Strongly Agree responses across all dimensions. Task Awareness and Thought Structuring show the largest differences (both marked p<.01), with \sysname{} responses concentrated in Agree to Strongly Agree. AI's Understanding, Authorship, Controllability, and Overall Satisfaction are also significantly higher for \sysname{} (all p<.05). Baseline responses are more dispersed, with notable portions in Neutral to Disagree ranges.}
    \label{fig:post-survey}
\end{figure*}
\section{Evaluation}
We conducted a within-subjects user study to understand how a canvas-based context representation and in-situ snippet memoing shape users' experience of collaborating with AI on their task, by comparing \sysname{} with a baseline condition. Specifically, we investigated the following research questions:
\begin{itemize}
    \item \textbf{RQ1}: Does the canvas-based representation in \sysname{} help users maintain clearer awareness and control over their evolving task context?
    \item \textbf{RQ2}: Do user-generated in-situ snippets improve the quality of AI assistance during tasks?
\end{itemize}
Participants completed tasks in two conditions, counterbalanced. In the \textbf{\sysname{}} condition, participants used the full system as described above (Section~\ref{section:contexty}). As the effect of in-situ snippet memoing was already explored in our design exploration (Section~\ref{section:design_explore}), we designed the baseline to isolate the effect of the canvas representation while retaining context capture. In the \textbf{Baseline} condition, the canvas was removed---participants could not view the underlying memory structure or the canvas overview tab. The timeline view, snippet memoing, computer-use observation, and context-augmented chat (including \texttt{@}-mention of memory items) remained available. This baseline approximates how users interact with existing LLM services (e.g., ChatGPT Atlas~\cite{chatgptAtlas}), extended with snippet memoing and passive context capture. This setup allows us to further assess the effect of snippet memoing during task sessions in both conditions, while isolating the additional contribution of the canvas.

\subsection{Participants and Tasks}
We recruited 12 participants (5 female, 7 male; age $M=26.0$, $SD=3.28$) through online recruitment postings on our university community platform. All participants had prior experience using LLMs (e.g., ChatGPT, Claude, Gemini) for exploratory or sensemaking tasks. Participants were compensated KRW 40,000 ($\approx$ USD 28) for approximately two hours of their time.

To ensure comparable levels of familiarity, motivation, and initial cognitive engagement across participants, we provided three categories of exploratory sensemaking tasks: career-related tasks (e.g., exploring PhD labs, searching for internships), academic and research-related tasks (e.g., surveying related work, organizing research directions), and everyday life tasks (e.g., trip planning, product comparison).
In a pre-study survey, participants rated their familiarity with each category on a 5-point Likert scale and selected the two categories they felt most familiar with. To keep the two sessions comparable, we included only participants whose familiarity ratings for the two selected categories differed by at most one point, so that neither condition was paired with a substantially more familiar task.
They then brought one real-world task from each selected category that they were genuinely motivated to pursue. 
Rather than assigning a fixed task, we adopted this open task design for ecological validity, so that participants worked on tasks of genuine personal relevance and brought their own criteria and prior knowledge to bear.

\subsection{Procedure}
Each session lasted approximately two hours. The session began with a \textbf{10-minute introduction}, during which the study was explained, and participants provided informed consent. Participants then completed \textbf{two task sessions} (45 minutes each), separated by a short break. In each session, participants first completed a 10-minute system walkthrough with a warm-up task to familiarize themselves with the interface, conducted the task work under the assigned condition for 30 minutes, and responded to a 5-minute post-condition survey. The session concluded with a \textbf{15-minute semi-structured interview} to explore participants' subjective experiences across both conditions.

\subsection{Measures}
\subsubsection{\textbf{Effect of the Canvas (RQ1)}}
To evaluate the effect of the canvas, we used a post-condition survey after each condition with 7-point Likert scales. The survey captured six dimensions related to user experience and perceived control: \textit{Task Awareness}, \textit{Thought Structuring}, \textit{Perceived AI's Understanding}, \textit{Authorship}, \textit{Controllability}, and \textit{Overall Satisfaction}. 
We then compared responses between conditions using Wilcoxon signed-rank tests.
The exact wording of all questionnaire items is provided in Appendix~\ref{app:questionnaire}.

\subsubsection{\textbf{Effect of Snippets (RQ2)}}
To assess whether in-situ snippet memoing improves AI response quality, we employed an \textbf{in-situ preference probe} during task sessions in both baseline and \sysname{} conditions. For each user query, the system silently generated two candidate responses in parallel: one using the full context, and one with all snippet-derived context filtered out. To avoid disrupting task flow, the rating was triggered only when the two contexts and their resulting responses were sufficiently different, determined by a two-stage gating mechanism based on context similarity and response-level divergence (see Appendix~\ref{appendix:gating_mechanism} for details). When triggered,
participants were shown the two responses in randomized order without knowing whether each response was generated with or without snippet-derived context. They selected their preferred response and could optionally provide a brief reason for their choice, which was further discussed during the post-task interview.
We report the selection ratio of snippet-informed responses as an indicator of snippet utility.
\section{Results}
\sysname{} significantly outperformed the baseline across all six dimensions (Figure~\ref{fig:post-survey}). Beyond these overall improvements, participants reported better task awareness, stronger authorship and control, and improved alignment with the AI. They also engaged with the canvas through both active refinement and passive inspection, while snippet memoing improved response quality and supported more intentional context sharing. We report the findings organized into themes based on quantitative and interview data.
\subsection{RQ1: Effect of the Canvas on Task Awareness and Sense of Control}

\subsubsection{\textbf{Enhancing Task Awareness and Reinforcing Authorship}}
The canvas significantly improved participants’ \textit{task awareness} ($M=5.67$ vs. $4.33$, $\Delta=+1.33$, $W=0.0$, $p=.0039^{**}$) and \textit{authorship} ($M=5.33$ vs. $4.33$, $\Delta=+1.00$, $W=2.0$, $p=.0312^{*}$). \sysname{} not only helped users maintain awareness of their ongoing task, but also supported a stronger sense that the task reflected their own thinking rather than the AI's suggestions. Participants described the canvas as making their task progress visible and traceable, enabling them to revisit prior steps without mentally reconstructing them (P01-8, P10, P12). Importantly, this visibility contributed to a sense of ownership. Participants (P02, P05, P12) described the canvas not as an AI-managed memory, but as their own space for maintaining and shaping task context. As P12 noted, \textit{``gave me a sense of ownership over the task, this in contrast to the conventional LLM systems where outputs often feel more like the AI’s product than my own.''}

\subsubsection{\textbf{Supporting Thought Structuring Through Both Active Organization and Passive Inspection}}
Participants reported significantly higher \textit{thought structuring} with \sysname{} ($M=5.83$ vs. $4.33$, $\Delta=+1.50$, $W=0.0$, $p=.0039^{**}$), indicating that the canvas effectively supported organization of information and ideas. Importantly, this effect emerged regardless of the usage patterns. Some participants (P02, P03, P05, P10-12) actively reorganized and edited canvas items, while others (P04, P06, P09) primarily relied on passively inspecting the AI-generated organization (Figure~\ref{fig:interaction-log} in the Appendix). These suggest two complementary pathways for supporting thought structuring: \textbf{\textit{active organization}} and \textbf{\textit{passive inspection}}.
Participants who engaged in \textit{\textbf{active reorganization}} treated the AI-generated organization as a starting point, editing and regrouping items to reflect their evolving understanding. P03 noted that reorganizing helped them identify what to explore next or what additional information to seek, while P12 described refining the structure as a way to clarify their own thinking as the task progressed.
In contrast, participants who engaged in \textit{\textbf{passive inspection}} used the canvas as a scaffold for understanding the current state of their task, glancing at the overview to reorient themselves and then returning to their task. P09 noted, ``I focused more on getting answers than reorganizing the canvas, yet when I briefly glanced at it, I felt my task context was fairly well structured.''

\subsubsection{\textbf{Fostering Trust and Alignment Through Visibility into AI's Understanding}}
Participants reported significantly higher \textit{perceived AI’s understanding} in \sysname{} than in the baseline ($M=5.25$ vs. $4.25$, $\Delta=+1.00$, $W=3.5$, $p=.0273^{*}$), suggesting that the canvas helped participants feel that the AI better understood their context and intentions during the task. This led to increased trust and alignment with the AI. By making the AI's interpretation of users' context visible and editable, users could inspect how their inputs were being interpreted and correct misalignments. P01 described that observing how the AI organized the information increased their trust. P10 noted that being able to see what the AI was referencing made it easier to verify and correct its understanding when needed.

\subsubsection{\textbf{Increased Controllability while Introducing a Tension with Curation Effort and Noise}}
\begin{figure}
    \centering
    \includegraphics[trim=0 30pt 0 90pt, clip, width=1\linewidth]{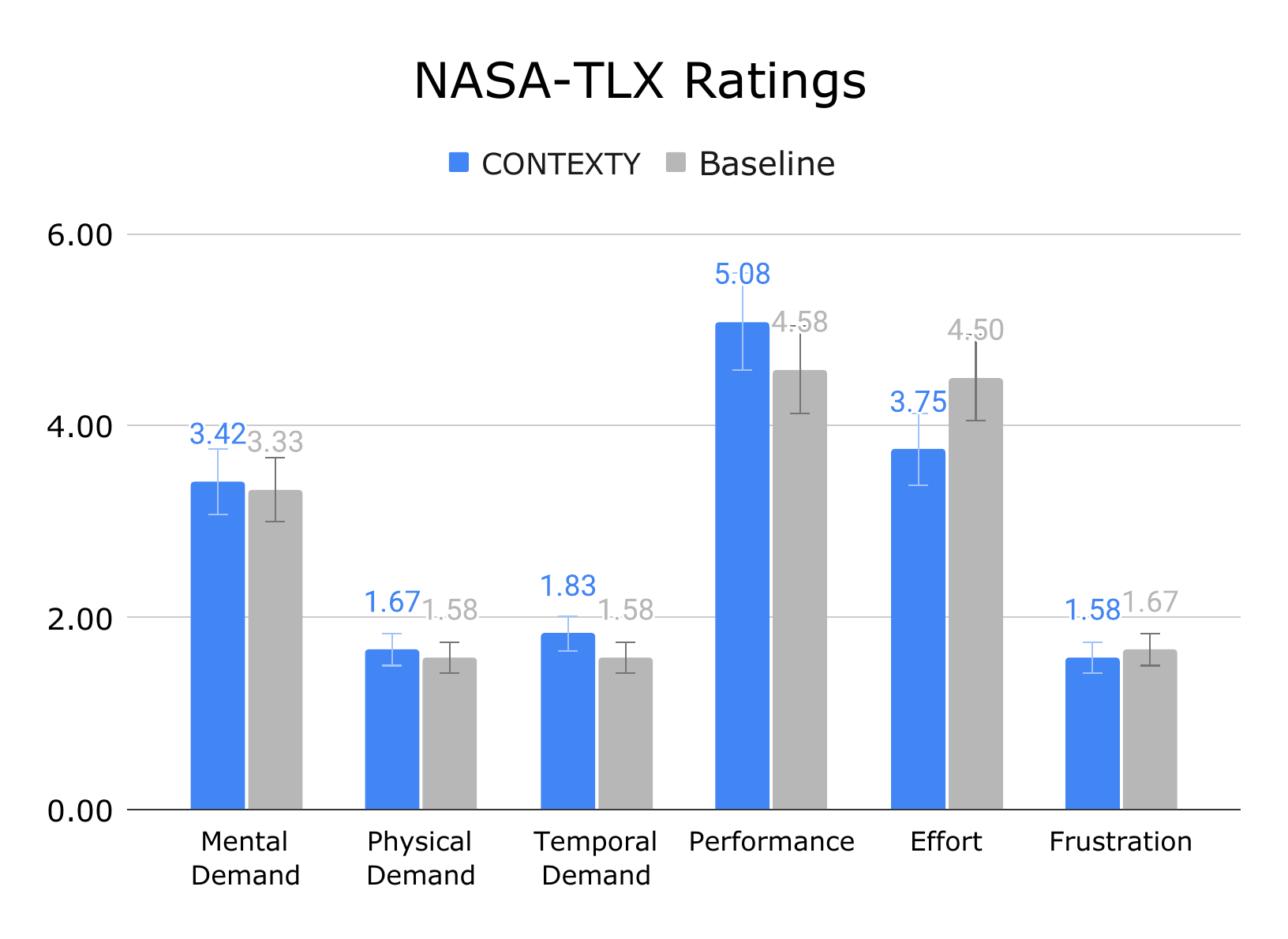}
    \caption{Comparison of NASA-TLX ratings between \sysname{} and Baseline.}
    \vspace{-10pt}
    \label{fig:nasa-tlx}
    \Description{This figure shows a grouped bar chart comparing NASA-TLX ratings between \sysname{} (blue bars) and the Baseline (gray bars) across six dimensions: Mental Demand, Physical Demand, Temporal Demand, Performance, Effort, and Frustration.}
\end{figure}
Participants reported significantly higher \textit{controllability} when using \sysname{} ($M=5.58$ vs. $4.00$, $\Delta=+1.58$, $W=6.0$, $p=.0332^{*}$), suggesting that the canvas allowed users to actively intervene in, modify, and curate their task context rather than relying solely on the AI’s automatic organization. However, qualitative findings reveal an important tension. While the canvas increased controllability, it also introduced the need for users to curate and filter the accumulated context. Because the system continuously captured information through automatic observation, participants often encountered irrelevant or low-value items, particularly in multitasking scenarios. Although it contributed to richer context capture, participants frequently reported that it introduced unnecessary information, like noise, which increases curation effort (P10, P12). P10 even expressed a preference not to use automatic observation to maintain a cleaner task context. 
Interestingly, this additional curation effort was not reflected in the NASA-TLX results, where no significant differences were observed (Figure~\ref{fig:nasa-tlx}). 
While \sysname{} introduces curation effort, the baseline required users to mentally organize and keep track of relationships among captured items using only a chronological list. The two conditions therefore differed in the form of effort rather than its overall amount, which may explain the comparable workload ratings.

\subsection{RQ2: Effect of In-situ Snippet Memoing on AI Response Quality}
\begin{figure}
    \centering
    \includegraphics[trim=0 10pt 0 5pt, clip, width=1\linewidth]{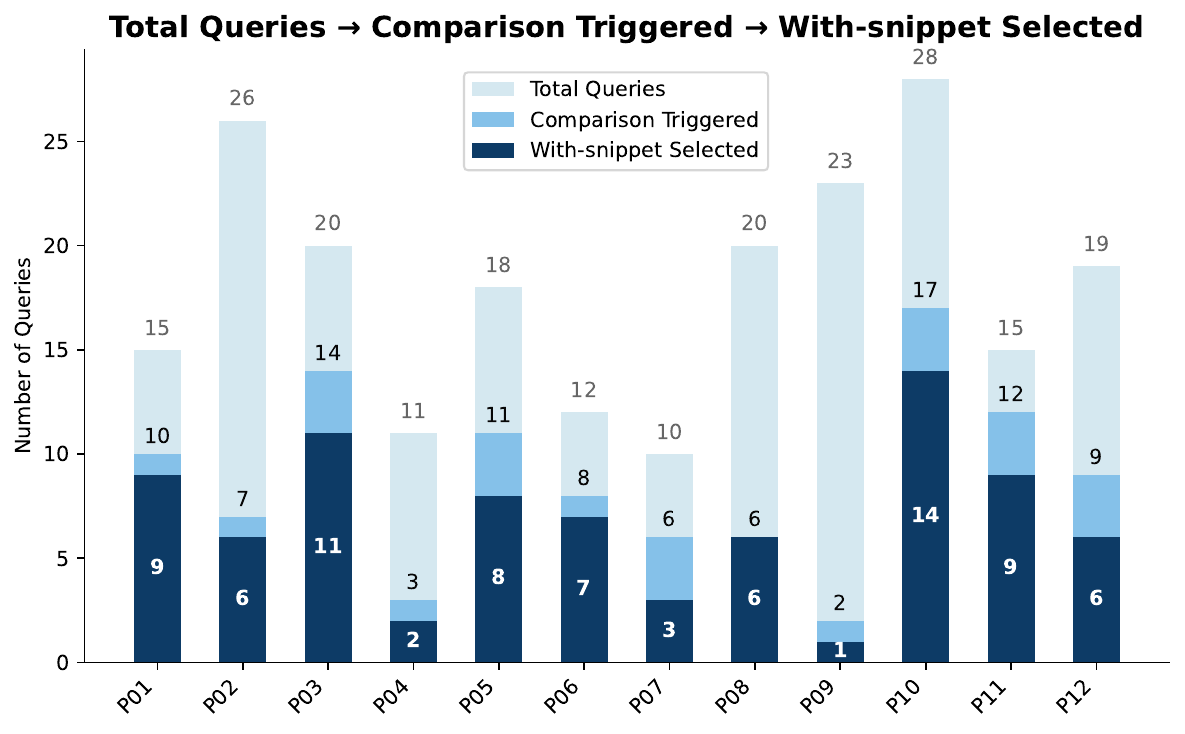}
    \caption{Per-participant query funnel (total queries, comparisons triggered, and with-snippet responses selected). On average, 48\% of queries triggered a comparison, and among those, 78\% resulted in selection of the with-snippet response.}
    \vspace{-10pt}
    \label{fig:snippet-funnel}
    \Description{A stacked bar chart showing three layers per participant (P01--P12): total queries (light blue), comparison triggered (medium blue), and with-snippet selected (dark blue).  On average, 48\% of queries triggered a comparison, and among those, with-snippet responses were selected 78\% of the time.}
\end{figure}
\subsubsection{\textbf{With-Snippet Responses Were Consistently Preferred}}
Across all 12 participants, with-snippet responses were preferred at a rate of 78.1\% on average (Figure~\ref{fig:snippet-funnel}). On average, 8.8 out of 18.1 total queries per participant triggered the comparison (48\%), and 6.8 of those resulted in selection of the with-snippet response. 
When explaining their choices, participants reported that the with-snippet responses felt more directly grounded in their own context and perspective. P02 noted that the with-snippet response was clearly better at directly answering what they had asked, and P03 described how the with-snippet response used vocabulary and framing that matched their own way of thinking about the task. 
However, the preference for with-snippet responses was not always the case. In some cases, participants favored responses generated without snippets as they provided more diverse perspectives or introduced ideas beyond their immediate context. This suggests that while snippets reliably improve response relevance, they may also narrow the response space, leading users to occasionally prefer less constrained answers, especially for ideation or exploration.

\subsubsection{\textbf{Snippet Memoing Integrates Naturally into Task Flow, and the Canvas Amplifies Its Effect}}
Interaction log analysis shows that snippet-memoing events occurred steadily throughout sessions (Figure~\ref{fig:interaction-log} in the Appendix), indicating that it naturally integrated into users' task flow. Participants stated that the snippet memoing, as a lightweight action, did not require switching context or interrupting ongoing work. Notably, several participants reported that the act of writing a snippet memo itself helped them think (P01, P03-9, P12). P01 described how capturing a snippet sometimes steered their thinking in a new direction, and P03 noted that the process of articulating why something mattered helped them clarify what to explore next.
Participants reported that their use of snippet memoing differed depending on the presence of the canvas. In the baseline, memos were often brief, summary-like labels primarily for later retrieval. In contrast, with \sysname{}, the canvas enabled these snippets to be organized, which led participants to record richer thoughts, including connections between pieces of information and personal interpretations (P03–4, P06–7, P11). Participants emphasized that snippets alone were not enough; the canvas gave them structure and made them retrievable and interpretable. As P03 noted, ``snippets let me express how I interpret the information, but the canvas is what made those interpretations visible and usable.''

\subsubsection{\textbf{Trade-offs Between Passive Computer-Use Observation and Snippet Memoing}}
While passive observation automatically captured background task context~\cite{shaikh2025GUM}, participants found it insufficient for conveying their intent and interpretation. As P05 noted, screen captures reflected what they were looking at, but not \textbf{why} it mattered or \textbf{how} they were making sense of it. In contrast, snippets allowed users to actively shape how their context was shared with the AI. Participants noted that this helped them communicate what they considered important (P03), such as connecting new information to prior knowledge or providing examples to clarify intent (P07, P10).
At the same time, participants acknowledged that continuous observation could capture broader context and even retain information they might otherwise miss. However, these benefits were often outweighed by concerns about noise and unintended information. 
Notably, no participant disabled passive observation during the study, though several remarked in the post-task interview that they would prefer limiting or turning off observation, noting that while they could tell when to turn it off for privacy, it was less clear what made it worth keeping on.
Together, these findings suggest that while observation provides useful background context, snippet memoing serves as a more intentional, user-controllable channel for sharing their task context that matters more for AI assistance.
\section{Discussion}
Building on our findings, we discuss (1) how systems can capture user context while minimizing user burden, (2) how such context can be represented in forms that are understandable by both users and AI, and (3) how such an approach can extend beyond single sessions to support sustainable context use over time.

Our findings reveal a tension between comprehensive context capture and selective control. Passive observation contributed to a richer context by capturing information users might otherwise miss, but also introduced noise and an additional curation burden, particularly in multitasking scenarios. At the same time, relying solely on user-driven externalization places a burden on the user. Balancing these two modalities is therefore an important design challenge. 
One promising approach to navigating this balance is to let the system learn from users' behavior. By leveraging snippet memoing patterns---what and where users choose to capture, and the thoughts they attach as memos---alongside patterns of context items actively referenced in chat, the system can infer which types of information tend to matter to an individual user. This inference could then drive proactive surfacing of relevant signals from observed behavior, such as highlighting passages on pages similar to previously snippeted content or flagging new information related to concepts the user frequently referenced in their memos.
However, as agents take a more active role in shaping and surfacing such context, there is a risk that AI-generated organizations may inadvertently steer how users understand and pursue their tasks. Ensuring that such support remains transparent and correctable by the user will be an important consideration for future work.

In human teamwork, effective collaboration depends on building and maintaining a \textit{shared mental model} of the task, including each other's goals, priorities, and evolving understanding~\cite{cannon1993shared, mathieu2000influence}. Our findings suggest a similar need in human–AI collaboration: making task context shared and inspectable improved users' sense of authorship, control, and perceived AI understanding of the task. Yet this raises an important open question: what should such a shared representation actually look like? Humans benefit from spatial, visual organization that supports reflection and navigation~\cite{clark1991dual, shipman1999spatial}. AI agents, by contrast, may need more structured textual representations that can be incorporated into model inputs, such as \texttt{llms.txt}~\cite{llmtext}, which encodes complex content in LLM-readable format. This suggests the need for a translation layer between human-facing and AI-facing representations of context. 

\sysname{} takes an initial step in this direction by externalizing the AI’s context on a canvas that users can inspect and revise. The canvas presents context in an intuitive, spatial form, while grounding it in the underlying hierarchical memory structure used by the AI. In this way, the canvas functions as a translation layer that makes AI context accessible to users, while also allowing users’ cognitive processes to be incorporated. This approach is facilitated by the form of AI context used in our system, which organizes LLM memories into a hierarchical structure inspired by recent LLM memory work~\cite{xu2025Amem, sun2026h}. However, as AI context representations evolve, they may not always align with formats that are easily interpretable or editable by users. Future work should therefore explore how such translation layers can generalize beyond specific representations, and how to maintain alignment between human-meaningful context and model-effective representations as both continue to evolve.

The synergy between in-situ snippet capturing and canvas-based organization also points to a broader design opportunity of embedding cognitive trail into everyday work environments. 
Recent AI-powered browsers have begun to integrate AI into users’ ongoing knowledge work, supporting activities such as information gathering and assistance within the browsing experience~\cite{comet, chatgptAtlas, dia, jiang2025orca}.
Our findings suggest that these systems could benefit from making users’ cognitive processes explicit and inspectable. Rather than leaving them implicitly inferred by AI or transient, a canvas-like layer could make users' cognitive trail visible and continuously usable within their primary work environment.

Finally, the notion of user-inspectable and correctable AI context, grounded in users’ cognitive traces, carries important implications beyond single-session interactions. As this context accumulates over time, it calls for more systematic approaches to managing and reusing memory. This includes distinguishing between short-term and long-term context~\cite{sumers2023cognitive, hatalis2023memory}, organizing information at the task or project level, and supporting selective reuse of relevant past context across sessions. This opens up opportunities for more sustainable interaction, where past cognitive traces can be maintained, refined, and brought forward when needed. To support this, future work should explore mechanisms for filtering and prioritizing accumulated context, helping users decide what to retain, revisit, or discard over time. This includes developing representations and interaction techniques that allow users to selectively surface relevant past context without being overwhelmed, while preserving their ability to inspect and correct how it is used by the AI.

\section{Limitations}
Our study focuses on single-session tasks, limiting our understanding of how context-sharing behaviors evolve over time; future work should examine how users manage and reuse accumulated context across sessions through longitudinal deployment. The tasks we studied primarily involved information foraging and sensemaking, leaving open how snippet usage and canvas-based organization may differ in artifact-producing tasks such as writing or programming. Finally, although \sysname{} externalizes AI context in a user-inspectable and correctable form, it is grounded in a hierarchical memory structure, which may limit its generalizability across tasks and longer-term use.

\section{Conclusion}
We presented \sysname{}, a system that incorporates users’ in-situ thoughts into AI context and makes it visible and editable in a shared space for human–AI collaboration. Our evaluation showed that \sysname{} improves users’ task awareness, sense of authorship, control, and perceived AI understanding.
Our work highlights the importance of moving beyond implicitly inferred, system-managed context toward user-inspectable and correctable AI context that incorporates users’ cognitive processes. We hope this work motivates future systems that support more transparent and collaborative ways of constructing context, keeping humans and AI aligned in complex tasks.
\begin{acks}
We appreciate the reviewers for their valuable feedback and comments. We also thank the members of KIXLAB for their help in polishing this manuscript.
This work was supported by the National Research Foundation of Korea (NRF) grant funded by the Korea government (Ministry of Science and ICT) (No.RS-2025-00557726).
This work was also supported by the Office of Naval Research (ONR: N00014-24-1-2290).
\end{acks}

\bibliographystyle{ACM-Reference-Format}
\bibliography{references}

@String{Computing = "Computing" }

@String{Computer = "{IEEE} Computer" }

@String{Springer = "Springer-Verlag" }

@article{Hollan2000Distributed,
author = {Hollan, James and Hutchins, Edwin and Kirsh, David},
title = {Distributed cognition: toward a new foundation for human-computer interaction research},
year = {2000},
issue_date = {June 2000},
publisher = {Association for Computing Machinery},
address = {New York, NY, USA},
volume = {7},
number = {2},
issn = {1073-0516},
url = {https://doi.org/10.1145/353485.353487},
doi = {10.1145/353485.353487},
journal = {ACM Trans. Comput.-Hum. Interact.},
month = jun,
pages = {174–196},
numpages = {23}
}

@inbook{Norman1991CognitiveArtifacts,
author = {Norman, Donald},
year = {1991},
month = {01},
pages = {333},
title = {Cognitive Artifacts},
isbn = {0521400562},
journal = {Designing interaction: psychology at the human-computer interface}
}

@inproceedings{shaikh2025GUM,
author = {Shaikh, Omar and Sapkota, Shardul and Rizvi, Shan and Horvitz, Eric and Park, Joon Sung and Yang, Diyi and Bernstein, Michael S.},
title = {Creating General User Models from Computer Use},
year = {2025},
isbn = {9798400720376},
publisher = {Association for Computing Machinery},
address = {New York, NY, USA},
url = {https://doi.org/10.1145/3746059.3747722},
doi = {10.1145/3746059.3747722},
booktitle = {Proceedings of the 38th Annual ACM Symposium on User Interface Software and Technology},
articleno = {35},
numpages = {23},
location = {
},
series = {UIST '25}
}

@inproceedings{tashman2011liquidtext,
author = {Tashman, Craig S. and Edwards, W. Keith},
title = {LiquidText: a flexible, multitouch environment to support active reading},
year = {2011},
isbn = {9781450302289},
publisher = {Association for Computing Machinery},
address = {New York, NY, USA},
url = {https://doi.org/10.1145/1978942.1979430},
doi = {10.1145/1978942.1979430},
booktitle = {Proceedings of the SIGCHI Conference on Human Factors in Computing Systems},
pages = {3285–3294},
numpages = {10},
location = {Vancouver, BC, Canada},
series = {CHI '11}
}

@misc{wang2025openhandsopenplatformai,
      title={OpenHands: An Open Platform for AI Software Developers as Generalist Agents}, 
      author={Xingyao Wang and Boxuan Li and Yufan Song and Frank F. Xu and Xiangru Tang and Mingchen Zhuge and Jiayi Pan and Yueqi Song and Bowen Li and Jaskirat Singh and Hoang H. Tran and Fuqiang Li and Ren Ma and Mingzhang Zheng and Bill Qian and Yanjun Shao and Niklas Muennighoff and Yizhe Zhang and Binyuan Hui and Junyang Lin and Robert Brennan and Hao Peng and Heng Ji and Graham Neubig},
      year={2025},
      eprint={2407.16741},
      archivePrefix={arXiv},
      primaryClass={cs.SE},
      url={https://arxiv.org/abs/2407.16741}, 
}

@misc{shaikh2026learningactionpredictorshumancomputer,
      title={Learning Next Action Predictors from Human-Computer Interaction}, 
      author={Omar Shaikh and Valentin Teutschbein and Kanishk Gandhi and Yikun Chi and Nick Haber and Thomas Robinson and Nilam Ram and Byron Reeves and Sherry Yang and Michael S. Bernstein and Diyi Yang},
      year={2026},
      eprint={2603.05923},
      archivePrefix={arXiv},
      primaryClass={cs.CL},
      url={https://arxiv.org/abs/2603.05923}, 
}

@article{zhang1997nature,
  title={The nature of external representations in problem solving},
  author={Zhang, Jiajie},
  journal={Cognitive science},
  volume={21},
  number={2},
  pages={179--217},
  year={1997},
  publisher={Elsevier}
}

@book{schon1992reflective,
  title={The reflective practitioner: How professionals think in action},
  author={Sch{\"o}n, Donald A},
  year={1992},
  publisher={Routledge},
  url = {https://doi.org/10.4324/9781315237473}
}

@book{ahrens2022take,
  title={How to take smart notes: One simple technique to boost writing, learning and thinking},
  author={Ahrens, S{\"o}nke},
  year={2022},
  publisher={S{\"o}nke Ahrens}
}

@article{xu2025Amem,
  title={A-mem: Agentic memory for llm agents},
  author={Xu, Wujiang and Liang, Zujie and Mei, Kai and Gao, Hang and Tan, Juntao and Zhang, Yongfeng},
  journal={arXiv preprint arXiv:2502.12110},
  year={2025}
}

@article{lee2025choir,
  title={CHOIR: A Chatbot-mediated Organizational Memory Leveraging Communication in University Research Labs},
  author={Lee, Sangwook and Abbas, Adnan and Chen, Yan and Kim, Young-Ho and Lee, Sang Won},
  journal={arXiv preprint arXiv:2509.20512},
  year={2025}
}

@inproceedings{sun2026h,
  title={H-MEM: Hierarchical Memory for High-Efficiency Long-Term Reasoning in LLM Agents},
  author={Sun, Haoran and Zeng, Shaoning and Zhang, Bob},
  booktitle={Proceedings of the 19th Conference of the European Chapter of the Association for Computational Linguistics (Volume 1: Long Papers)},
  pages={341--350},
  year={2026}
}

@article{gmeiner2026pointaloud,
  title={PointAloud: An Interaction Suite for AI-Supported Pointer-Centric Think-Aloud Computing},
  author={Gmeiner, Frederic and Thompson, John and Fitzmaurice, George and Matejka, Justin},
  journal={arXiv preprint arXiv:2602.09296},
  year={2026}
}

@article{son2025clearfairy,
  title={ClearFairy: Capturing Creative Workflows through Decision Structuring, In-Situ Questioning, and Rationale Inference},
  author={Son, Kihoon and Choi, DaEun and Kim, Tae Soo and Kim, Young-Ho and Yun, Sangdoo and Kim, Juho},
  journal={arXiv preprint arXiv:2509.14537},
  year={2025}
}

@article{li2026orality,
  title={Orality: A Semantic Canvas for Externalizing and Clarifying Thoughts with Speech},
  author={Li, Wengxi and Tian, Jingze and Liu, Can},
  journal={arXiv preprint arXiv:2603.02544},
  year={2026}
}

@inproceedings{suh2023sensecape,
  title={Sensecape: Enabling multilevel exploration and sensemaking with large language models},
  author={Suh, Sangho and Min, Bryan and Palani, Srishti and Xia, Haijun},
  booktitle={Proceedings of the 36th annual ACM symposium on user interface software and technology},
  pages={1--18},
  year={2023}
}

@inproceedings{jiang2023graphologue,
  title={Graphologue: Exploring large language model responses with interactive diagrams},
  author={Jiang, Peiling and Rayan, Jude and Dow, Steven P and Xia, Haijun},
  booktitle={Proceedings of the 36th annual ACM symposium on user interface software and technology},
  pages={1--20},
  year={2023}
}

@inproceedings{suh2024luminate,
  title={Luminate: Structured generation and exploration of design space with large language models for human-ai co-creation},
  author={Suh, Sangho and Chen, Meng and Min, Bryan and Li, Toby Jia-Jun and Xia, Haijun},
  booktitle={Proceedings of the 2024 CHI Conference on Human Factors in Computing Systems},
  pages={1--26},
  year={2024}
}

@inproceedings{krosnick2021think,
  title={Think-aloud computing: Supporting rich and low-effort knowledge capture},
  author={Krosnick, Rebecca and Anderson, Fraser and Matejka, Justin and Oney, Steve and S. Lasecki, Walter and Grossman, Tovi and Fitzmaurice, George},
  booktitle={Proceedings of the 2021 CHI conference on human factors in computing systems},
  pages={1--13},
  year={2021}
}

@inproceedings{kuznetsov2022fuse,
  title={Fuse: In-Situ sensemaking support in the browser},
  author={Kuznetsov, Andrew and Chang, Joseph Chee and Hahn, Nathan and Rachatasumrit, Napol and Breneisen, Bradley and Coupland, Julina and Kittur, Aniket},
  booktitle={Proceedings of the 35th Annual ACM Symposium on User Interface Software and Technology},
  pages={1--15},
  year={2022}
}

@inproceedings{marshall1997annotation,
  title={Annotation: from paper books to the digital library},
  author={Marshall, Catherine C},
  booktitle={Proceedings of the second ACM international conference on Digital libraries},
  pages={131--140},
  year={1997}
}

@inproceedings{zhang2025ladica,
  title={LADICA: a large shared display interface for generative AI cognitive assistance in co-located team collaboration},
  author={Zhang, Zheng and Peng, Weirui and Chen, Xinyue and Cao, Luke and Li, Toby Jia-Jun},
  booktitle={Proceedings of the 2025 CHI Conference on Human Factors in Computing Systems},
  pages={1--22},
  year={2025}
}

@inproceedings{lam2025policy,
  title={Policy Maps: Tools for Guiding the Unbounded Space of LLM Behaviors},
  author={Lam, Michelle S and Hohman, Fred and Moritz, Dominik and Bigham, Jeffrey P and Holstein, Kenneth and Kery, Mary Beth},
  booktitle={Proceedings of the 38th Annual ACM Symposium on User Interface Software and Technology},
  pages={1--24},
  year={2025}
}

@inproceedings{zhang2021conceptscope,
  title={Conceptscope: Organizing and visualizing knowledge in documents based on domain ontology},
  author={Zhang, Xiaoyu and Chandrasegaran, Senthil and Ma, Kwan-Liu},
  booktitle={Proceedings of the 2021 chi conference on human factors in computing systems},
  pages={1--13},
  year={2021}
}

@inproceedings{zhang2023concepteva,
  title={ConceptEVA: Concept-based interactive exploration and customization of document summaries},
  author={Zhang, Xiaoyu and Li, Jianping and Chi, Po-Wei and Chandrasegaran, Senthil and Ma, Kwan-Liu},
  booktitle={Proceedings of the 2023 CHI Conference on Human Factors in Computing Systems},
  pages={1--16},
  year={2023}
}

@inproceedings{zhao2025knoll,
  title={Knoll: Creating a knowledge ecosystem for large language models},
  author={Zhao, Dora and Yang, Diyi and Bernstein, Michael S},
  booktitle={Proceedings of the 38th Annual ACM Symposium on User Interface Software and Technology},
  pages={1--23},
  year={2025}
}

@article{mei2025survey,
  title={A survey of context engineering for large language models},
  author={Mei, Lingrui and Yao, Jiayu and Ge, Yuyao and Wang, Yiwei and Bi, Baolong and Cai, Yujun and Liu, Jiazhi and Li, Mingyu and Li, Zhong-Zhi and Zhang, Duzhen and others},
  journal={arXiv preprint arXiv:2507.13334},
  year={2025}
}

@article{yang2025contextagent,
  title={Contextagent: Context-aware proactive llm agents with open-world sensory perceptions},
  author={Yang, Bufang and Xu, Lilin and Zeng, Liekang and Liu, Kaiwei and Jiang, Siyang and Lu, Wenrui and Chen, Hongkai and Jiang, Xiaofan and Xing, Guoliang and Yan, Zhenyu},
  journal={arXiv preprint arXiv:2505.14668},
  year={2025}
}

@article{lu2024proactive,
  title={Proactive agent: Shifting llm agents from reactive responses to active assistance},
  author={Lu, Yaxi and Yang, Shenzhi and Qian, Cheng and Chen, Guirong and Luo, Qinyu and Wu, Yesai and Wang, Huadong and Cong, Xin and Zhang, Zhong and Lin, Yankai and others},
  journal={arXiv preprint arXiv:2410.12361},
  year={2024}
}

@inproceedings{gmeiner2025intenttagging,
author = {Gmeiner, Frederic and Marquardt, Nicolai and Bentley, Michael and Romat, Hugo and Pahud, Michel and Brown, David and Roseway, Asta and Martelaro, Nikolas and Holstein, Kenneth and Hinckley, Ken and Riche, Nathalie},
title = {Intent Tagging: Exploring Micro-Prompting Interactions for Supporting Granular Human-GenAI Co-Creation Workflows},
year = {2025},
isbn = {9798400713941},
publisher = {Association for Computing Machinery},
address = {New York, NY, USA},
url = {https://doi.org/10.1145/3706598.3713861},
doi = {10.1145/3706598.3713861},
booktitle = {Proceedings of the 2025 CHI Conference on Human Factors in Computing Systems},
articleno = {531},
numpages = {31},
location = {
},
series = {CHI '25}
}

@inproceedings{brade2023promptify,
  title={Promptify: Text-to-image generation through interactive prompt exploration with large language models},
  author={Brade, Stephen and Wang, Bryan and Sousa, Mauricio and Oore, Sageev and Grossman, Tovi},
  booktitle={Proceedings of the 36th Annual ACM Symposium on User Interface Software and Technology},
  pages={1--14},
  year={2023}
}

@inproceedings{kim2025intentflow,
author = {Kim, Yoonsu and Son, Kihoon and Kim, Seoyoung and Chin, Brandon and Kim, Juho},
title = {IntentFlow: Investigating Fluid Dynamics of Intent Communication in Generative AI},
year = {2026},
isbn = {9798400725630},
publisher = {Association for Computing Machinery},
address = {New York, NY, USA},
url = {https://doi.org/10.1145/3800645.3812999},
doi = {10.1145/3800645.3812999},
booktitle = {Proceedings of the 2026 Designing Interactive Systems Conference},
pages = {3804–3837},
numpages = {34},
location = {Singapore, Singapore},
series = {DIS '26}
}

@inproceedings{Wang2024PromptCharm,
author = {Wang, Zhijie and Huang, Yuheng and Song, Da and Ma, Lei and Zhang, Tianyi},
title = {PromptCharm: Text-to-Image Generation through Multi-modal Prompting and Refinement},
year = {2024},
isbn = {9798400703300},
publisher = {Association for Computing Machinery},
address = {New York, NY, USA},
url = {https://doi.org/10.1145/3613904.3642803},
doi = {10.1145/3613904.3642803},
booktitle = {Proceedings of the 2024 CHI Conference on Human Factors in Computing Systems},
articleno = {185},
numpages = {21},
location = {Honolulu, HI, USA},
series = {CHI '24}
}

@article{shneiderman1983direct,
  title={Direct manipulation: A step beyond programming languages},
  author={Shneiderman, Ben},
  journal={Computer},
  volume={16},
  number={08},
  pages={57--69},
  year={1983},
  publisher={IEEE Computer Society}
}

@article{lam2025just,
  title={Just-In-Time Objectives: A General Approach for Specialized AI Interactions},
  author={Lam, Michelle S and Shaikh, Omar and Xu, Hallie and Guo, Alice and Yang, Diyi and Heer, Jeffrey and Landay, James A and Bernstein, Michael S},
  journal={arXiv preprint arXiv:2510.14591},
  year={2025}
}

@misc{tldraw,
author = {tldraw},
  title = {tldraw: Infinite Canvas SDK for React},
  url = {https://tldraw.dev/},
  note = {"[Accessed 2026-03-31]"}
}

@misc{electron,
author = {Electron},
  title = {Build cross-platform desktop apps with JavaScript, HTML, and CSS | Electron},
  url = {https://www.electronjs.org/},
  note = {"[Accessed 2026-03-31]"}
}

@inproceedings{yang26guide,
    title={GUIDE: A Benchmark for Understanding and Assisting Users in Open-Ended GUI Tasks}, 
    author={Saelyne Yang and Jaesang Yu and Yi-Hao Peng and Kevin Qinghong Lin and Jae Won Cho and Yale Song and Juho Kim},
    booktitle = {{Proceedings of the IEEE/CVF Conference on Computer Vision and Pattern Recognition (CVPR)}},
  year      = {2026},
  note      = {To appear},
}

@article{jiang2025orca,
  title={Orca: Browsing at scale through user-driven and ai-facilitated orchestration across malleable webpages},
  author={Jiang, Peiling and Xia, Haijun},
  journal={arXiv preprint arXiv:2505.22831},
  year={2025}
}

@article{mathieu2000influence,
  title={The influence of shared mental models on team process and performance.},
  author={Mathieu, John E and Heffner, Tonia S and Goodwin, Gerald F and Salas, Eduardo and Cannon-Bowers, Janis A},
  journal={Journal of applied psychology},
  volume={85},
  number={2},
  pages={273},
  year={2000},
  publisher={American Psychological Association}
}

@article{cannon1993shared,
  title={Shared mental models in expert team decision making.},
  author={Cannon-Bowers, Janis A and Salas, Eduardo and Converse, Sharolyn},
  year={1993},
  publisher={Lawrence Erlbaum Associates, Inc}
}

@misc{comet,
author={Perplexity},
  title = {Comet Browser: a Personal AI Assistant},
  url = {https://www.perplexity.ai/comet},
  note = {[Accessed 2026-03-31]}
}

@misc{chatgptAtlas,
author = {OpenAI},
  title = {ChatGPT Atlas},
  url = {https://chatgpt.com/atlas/},
  note = {[Accessed 2026-03-31]}
}

@misc{dia,
author = {Dia},
  title = {Dia Browser | AI Chat With Your Tabs},
  url = {https://www.diabrowser.com/},
  note = {[Accessed 2026-03-31]}
}

@misc{llmtext,
author = {Jeremy Howard},
  title = {The /llms.txt file – llms-txt},
  url = {https://llmstxt.org/},
  note = {[Accessed 2026-03-31]}
}

@misc{chatgptmemory,
author = {OpenAI},
  title = {What is Memory? - OpenAI Help Center},
  url = {https://help.openai.com/en/articles/8983136-what-is-memory},
  month = {},
  year = {2024},
  note = {[Accessed 2026-03-31]}
}

@misc{claudecode,
author = {Anthropic},
  title = {How Claude Code works - Claude Code Docs},
  url = {https://code.claude.com/docs/en/how-claude-code-works},
  month = {},
  year = {2025},
  note = {[Accessed 2026-03-31]}
}

@inproceedings{yang2006block,
  title={Block mean value based image perceptual hashing},
  author={Yang, Bian and Gu, Fan and Niu, Xiamu},
  booktitle={2006 International Conference on Intelligent Information Hiding and Multimedia},
  pages={167--172},
  year={2006},
  organization={IEEE}
}

@article{sumers2023cognitive,
  title={Cognitive architectures for language agents},
  author={Sumers, Theodore and Yao, Shunyu and Narasimhan, Karthik R and Griffiths, Thomas L},
  journal={Transactions on Machine Learning Research},
  year={2023}
}

@inproceedings{hatalis2023memory,
  title={Memory matters: The need to improve long-term memory in llm-agents},
  author={Hatalis, Kostas and Christou, Despina and Myers, Joshua and Jones, Steven and Lambert, Keith and Amos-Binks, Adam and Dannenhauer, Zohreh and Dannenhauer, Dustin},
  booktitle={Proceedings of the AAAI Symposium Series},
  volume={2},
  number={1},
  pages={277--280},
  year={2023}
}

@article{clark1991dual,
  title={Dual coding theory and education},
  author={Clark, James M and Paivio, Allan},
  journal={Educational psychology review},
  volume={3},
  number={3},
  pages={149--210},
  year={1991},
  publisher={Springer}
}

@article{shipman1999spatial,
  title={Spatial hypertext: an alternative to navigational and semantic links},
  author={Shipman III, Frank M and Marshall, Catherine C},
  journal={ACM Computing Surveys (CSUR)},
  volume={31},
  number={4es},
  pages={14--es},
  year={1999},
  publisher={ACM New York, NY, USA}
}

@inproceedings{yen2024memolet,
author = {Yen, Ryan and Zhao, Jian},
title = {Memolet: Reifying the Reuse of User-AI Conversational Memories},
year = {2024},
isbn = {9798400706288},
publisher = {Association for Computing Machinery},
address = {New York, NY, USA},
url = {https://doi.org/10.1145/3654777.3676388},
doi = {10.1145/3654777.3676388},
booktitle = {Proceedings of the 37th Annual ACM Symposium on User Interface Software and Technology},
articleno = {58},
numpages = {22},
location = {Pittsburgh, PA, USA},
series = {UIST '24}
}

@inproceedings{vaithilingam2025semantic,
author = {Vaithilingam, Priyan and Kim, Munyeong and Acosta-Parenteau, Frida-Cecilia and Lee, Daniel and Mhedhbi, Amine and Glassman, Elena L. and Arawjo, Ian},
title = {Semantic Commit: Helping Users Update Intent Specifications for AI Memory at Scale},
year = {2025},
isbn = {9798400720376},
publisher = {Association for Computing Machinery},
address = {New York, NY, USA},
url = {https://doi.org/10.1145/3746059.3747778},
doi = {10.1145/3746059.3747778},
booktitle = {Proceedings of the 38th Annual ACM Symposium on User Interface Software and Technology},
articleno = {137},
numpages = {18},
location = {
},
series = {UIST '25}
}

@inproceedings{kang2023Synergi,
author = {Kang, Hyeonsu B and Wu, Tongshuang and Chang, Joseph Chee and Kittur, Aniket},
title = {Synergi: A Mixed-Initiative System for Scholarly Synthesis and Sensemaking},
year = {2023},
isbn = {9798400701320},
publisher = {Association for Computing Machinery},
address = {New York, NY, USA},
url = {https://doi.org/10.1145/3586183.3606759},
doi = {10.1145/3586183.3606759},
booktitle = {Proceedings of the 36th Annual ACM Symposium on User Interface Software and Technology},
articleno = {43},
numpages = {19},
location = {San Francisco, CA, USA},
series = {UIST '23}
}

@inproceedings{ye2025ScholarMate,
author = {Ye, Runlong and Lee, Patrick and Varona, Matthew and Huang, Oliver and Nobre, Carolina},
title = {ScholarMate: A Mixed-Initiative Tool for Qualitative Knowledge Work and Information Sensemaking},
year = {2025},
isbn = {9798400713972},
publisher = {Association for Computing Machinery},
address = {New York, NY, USA},
url = {https://doi.org/10.1145/3707640.3731913},
doi = {10.1145/3707640.3731913},
booktitle = {Adjunct Proceedings of the 4th Annual Symposium on Human-Computer Interaction for Work},
articleno = {7},
numpages = {7},
location = {
},
series = {CHIWORK '25 Adjunct}
}
\clearpage
\appendix

\section{Context Retrieval for Chat}\label{appendix:retrieval}
When a user sends a query $q$, the system retrieves relevant memories in two stages.
\paragraph{Explicit References.} Any memories or branches explicitly referenced by the user (via @-mentions or ``Add to Chat'') are resolved first. For branch references, the system collects the branch's member memories sorted by recency and selects the most recent items as representatives.

\paragraph{Automatic Retrieval.}
The remaining non-archived memories are scored against the query using a composite ranking function:
\begin{align*}
\text{score}(m, q) &= \underbrace{\frac{|\text{tokens}(q) \cap \text{tokens}(m)|}{|\text{tokens}(q)|}}_{\text{token overlap}} \\
&+ \underbrace{0.2 \cdot \mathbf{1}[\exists\, t \in \text{tags}(m) : t \subseteq q]}_{\text{tag boost}} \\
&+ \underbrace{0.15 \cdot \max\!\Big(0,\; 1 - \tfrac{\text{age}(m)}{30\text{d}}\Big)}_{\text{recency}}
\end{align*} 
Here, $\text{tokens}(m)$ is derived from the concatenation of the memory's title, description, context, intent annotations, raw text, tags, and provenance metadata (application name, window title, URL). The tag boost rewards memories whose tags appear as substrings in the query, and the recency term decays linearly over a 30-day window.

\paragraph{Merging and Formatting.}
Explicitly referenced memories take priority; automatically retrieved memories fill the remaining slots. Each selected memory is formatted as a structured text block containing its source type, title, context, summary, and tags, and prepended to the user's message. Explicitly referenced items are flagged as \texttt{MENTION} to signal higher priority to the LLM.

\section{Observation Filtering Pipeline}\label{appendix:observation-filtering}
Passive observation can produce a high volume of captures, many of which are visually or semantically redundant. \sysname{} applies a two-stage filtering pipeline to manage this.

\paragraph{Stage 1: Perceptual Hash Deduplication.}
Before any LLM analysis, each observation screenshot is converted to a perceptual hash using the BMVBHash algorithm~\cite{yang2006block}. The image is resized to $16 \times 16$ grayscale and hashed into a 256-bit vector. The system computes the Hamming distance $d_H$ between the new hash and all previously accepted hashes:
$$d_H(h_{\text{new}}, h_i) = \sum_{j=1}^{256} \mathbf{1}[h_{\text{new}}^{(j)} \neq h_i^{(j)}]$$
If $d_H \leq 10$ for any existing hash, the screenshot is considered a near-duplicate and discarded before further processing. This lightweight check eliminates visually identical or near-identical captures (e.g., repeated screenshots of the same static page) without incurring LLM costs.

\paragraph{Stage 2: Semantic Similarity Auto-Hide.}
Screenshots that pass the perceptual hash check proceed through the standard analysis pipeline and are added to the memory tree. The system then evaluates whether the new observation is semantically redundant with respect to currently visible memories. Using the related-item retrieval mechanism, the system obtains the top-$k$ related items among the 30 most recent memories, each with an LLM-generated relevance score $\sigma_i \in [0,1]$. 
Let $\sigma^* = \max_i \sigma_i$ be the highest score among non-archived, non-hidden memories with source type \textit{observation} or \textit{snippet}. If $\sigma^* \geq 0.65$, the new observation is marked as hidden on the canvas. The memory is retained in the tree and remains available for context retrieval, but is not rendered on the canvas to reduce visual clutter.

This two-stage design is intentional: Stage 1 operates at the pixel level with negligible latency, filtering exact or near-exact recaptures. Stage 2 operates at the semantic level via LLM, catching cases where a user returns to the same task or topic in a different visual state (e.g., scrolling further on the same documentation page). Together, the two stages ensure that the canvas remains manageable while preserving the full capture history for retrieval.

\section{Preference Probe Implementation Details}\label{appendix:gating_mechanism}
\begin{figure*}
\centering
\includegraphics[width=\linewidth]{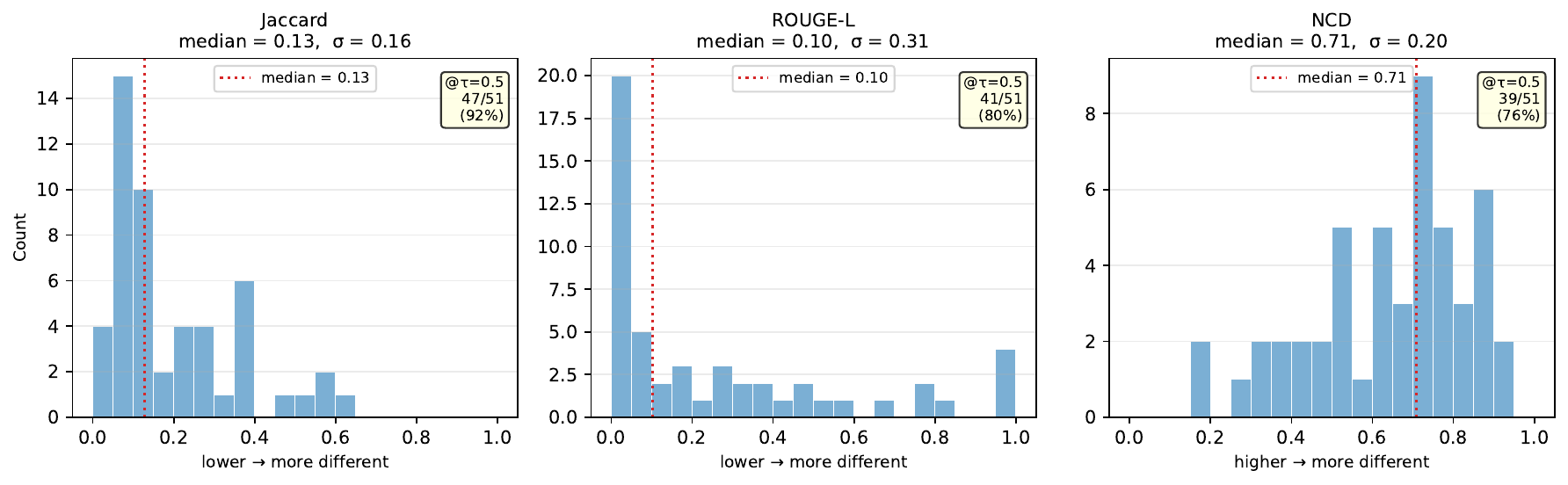}
\caption{Distribution of three similarity metrics across 51 response pairs from design exploration sessions. Dashed red lines indicate medians.}
\Description{Three histograms showing the distribution of Jaccard, ROUGE-L, and NCD similarity scores across 51 response pairs. Jaccard scores are concentrated near zero with a median of 0.13 and standard deviation of 0.16, with 92\% of pairs below the 0.5 threshold. ROUGE-L scores are also concentrated near zero with a median of 0.10 and standard deviation of 0.31, with 80\% below 0.5. NCD scores are concentrated near the upper end with a median of 0.71 and standard deviation of 0.20, with 76\% above 0.5. All three metrics indicate that most response pairs generated with and without snippet context differ substantially.}
\label{fig:metric_distributions}
\end{figure*}

\begin{figure*}[h]
\centering
\includegraphics[width=\linewidth]{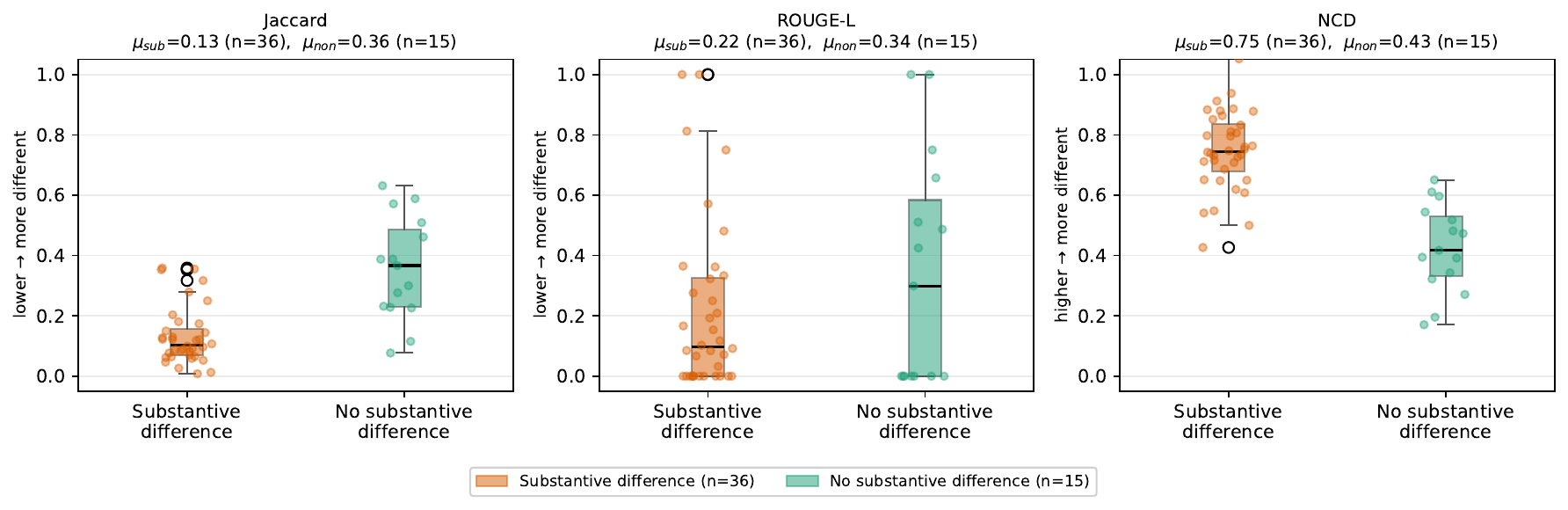}
\caption{Similarity metrics split by whether an independent LLM judge classified the response pair as exhibiting a substantive semantic difference. NCD provides the clearest separation between the two groups.}
\Description{Three box-and-jitter plots comparing Jaccard, ROUGE-L, and NCD scores between response pairs judged as substantively different (n=36, shown in orange) and not substantively different (n=15, shown in teal). For Jaccard, the substantive group has a mean of 0.13 and the non-substantive group 0.36, with overlapping distributions. For ROUGE-L, the means are 0.22 and 0.34, also with substantial overlap. For NCD, the substantive group has a mean of 0.75 and the non-substantive group 0.43, showing the clearest separation with minimal overlap between the two distributions. This demonstrates that NCD is the most discriminative metric for identifying substantively different response pairs.}
\label{fig:metrics_by_diff}
\end{figure*}
\paragraph{Context Construction.}
For each user query $q$, the system constructed two context variants in parallel: context $C_A$, built from the full memory tree (chat history + observations + snippets), and context $C_B$, built from the same tree but with all snippet-sourced memories and their associated intent annotations filtered out. Each context was independently scored and ranked using the retrieval function described in \autoref{appendix:retrieval}, ensuring that both variants selected their respective top memories under the same ranking criterion.

\paragraph{Stage 1: Context-Level Gate.}
To filter out cases where the two context variants were effectively equivalent, the system compared $C_A$ and $C_B$ using memory-level Jaccard similarity $J_{\text{mem}}$ (over selected memory IDs) and token-level Jaccard similarity $J_{\text{tok}}$ (over context text tokens). If $J_{\text{mem}} \geq 0.85$ and $J_{\text{tok}} \geq 0.92$, the contexts were deemed equivalent and only response $r_A$ was shown to the participant.

\paragraph{Stage 2: Response-Level Gate.}
When the contexts passed Stage 1, both were sent to the same LLM to generate candidate responses $r_A$ and $r_B$ in parallel. The system then applied a final gate based on Normalized Compression Distance (NCD) to determine whether the two responses were substantively different:
$$\text{NCD}(r_A, r_B) = \frac{|r_A \oplus r_B|_z - \min(|r_A|_z, |r_B|_z)}{\max(|r_A|_z, |r_B|_z)}$$ where $|\cdot|_z$ denotes the gzip-compressed byte length and $\oplus$ denotes string concatenation. Only when $\text{NCD}(r_A, r_B) > \tau$ were both responses presented to the participant.

\paragraph{Threshold Calibration.}
We set $\tau = 0.7$ based on a calibration analysis conducted on 51 response pairs collected from 10 participants during design exploration sessions (Section~\ref{section:design_explore}) prior to the main study. For each pair, we computed three similarity metrics---NCD, ROUGE-L, and Jaccard---and obtained an independent LLM judgment of whether the pair exhibited a substantive semantic difference. Figure~\ref{fig:metric_distributions} shows the distribution of each metric across all pairs. To assess discriminative power, Figure~\ref{fig:metrics_by_diff} splits each metric by the substantive-difference judgment. NCD exhibited the strongest separation ($\mu_{\text{sub}}$ = 0.75 vs.\ $\mu_{\text{non}}$ = 0.43), while Jaccard (0.13 vs.\ 0.36) and ROUGE-L (0.22 vs.\ 0.34) showed smaller gaps. Based on this, we adopted NCD with $\tau = 0.7$ as the operating point that best balanced probe sensitivity against participant disruption.

\begin{figure*}
    \centering
    \includegraphics[width=1\linewidth]{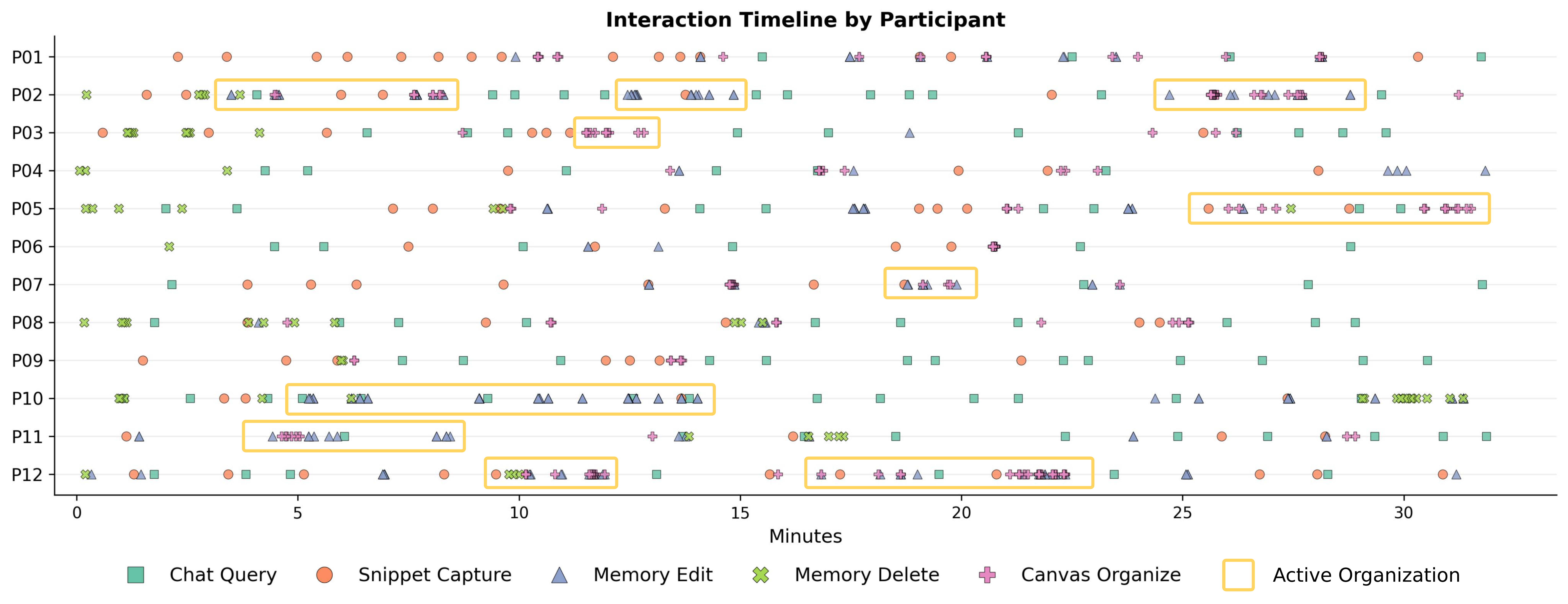}
    \caption{Interaction Log by Participant}
    \Description{A scatter-style timeline showing interaction events for 12 participants (P01--P12) over approximately 30 minutes of task sessions. Each row represents one participant, with the horizontal axis showing elapsed time in minutes. Six event types are plotted with distinct markers: Chat Query (green squares), Snippet Capture (orange circles), Memory Edit (purple triangles), Memory Delete (green crosses), Canvas Organize (pink diamonds), and Active Organization episodes (yellow rectangles spanning temporal ranges). Most participants show intermixed chat queries and snippet captures throughout their sessions, with memory edits scattered across the timeline. Active organization episodes, highlighted by yellow bounding boxes, occur in distinct bursts---for example, P02 organizes early (around minutes 3--7) and again later (minutes 25--30), while P10 has an extended organization phase spanning minutes 4--15. Participant activity density and interaction patterns vary considerably: some (e.g., P05, P06) show dense, continuous activity, while others (e.g., P04, P09) are sparser. Canvas organize and memory delete events are relatively infrequent compared to chat queries and snippet captures.}
    \label{fig:interaction-log}
\end{figure*}

\section{Post-Condition Questionnaire} \label{app:questionnaire}

After each condition, we administered a post-condition survey using 7-point Likert scales
(1 = strongly disagree, 7 = strongly agree). Table~\ref{tab:post_condition_questionnaire} lists all questionnaire items.

\begin{table}[H]
\centering
\small
\begin{tabular}{p{0.28\linewidth} p{0.64\linewidth}}
\toprule
\textbf{Dimension} & \textbf{Item} \\
\midrule
Task Awareness & ``The system helped me maintain an overview of the overall context of my task.'' \\

Thought Structuring & ``The system helped me organize and structure information and thoughts related to my task.'' \\

Perceived AI's Understanding & ``The AI seemed to understand my task context and intentions during the task.'' \\

Authorship & ``The system helped the task reflect my own thinking and judgment, not just the AI's suggestions.'' \\

Controllability & ``The system gave me control over how my task context was organized and represented.'' \\

Overall Satisfaction & ``I was satisfied with the overall interaction with the system.'' \\
\bottomrule
\end{tabular}
\Description{A two-column table presenting post-condition questionnaire items. Six dimensions are listed in the first column---Task Awareness, Thought Structuring, Perceived AI's Understanding, Authorship, Controllability, and Overall Satisfaction. Each is paired with a specific survey question item in quotes.}
\caption{Post-condition questionnaire items.}
\label{tab:post_condition_questionnaire}
\end{table}
\end{document}